\documentclass[aps,prd,onecolumn,12pt,amssymb,showpacs,superscriptaddress,nofootinbib,longbibliography]{revtex4-1}

\usepackage{amsfonts}
\usepackage{latexsym}
\usepackage{yfonts}
\usepackage{amsmath}
\usepackage{amssymb}
\usepackage{amsthm}
\usepackage{mathrsfs}
\usepackage{upgreek}
\usepackage{bm}
\usepackage{dcolumn}
\usepackage{epsfig}
\usepackage{graphicx}
\usepackage{latexsym}
\usepackage{rotating}
\usepackage{color}
\usepackage[dvipsnames]{xcolor}
\usepackage{soul}
\usepackage[english]{babel}
\usepackage[centering,hmargin=1in,vmargin=1in]{geometry}
\soulregister\cite7
\soulregister\ref7
\soulregister\pageref7

\usepackage{lmodern}
\usepackage[T1]{fontenc}
\usepackage[utf8]{inputenc}
\def\be{\begin{equation}}
\def\ee{\end{equation}}
\def\bea{\begin{eqnarray}}
\def\eea{\end{eqnarray}}
\usepackage{times}
\usepackage{setspace}

\usepackage{color}
\usepackage[unicode=true,pdfusetitle,
 bookmarks=true,bookmarksnumbered=false,bookmarksopen=false,
 breaklinks=false,pdfborder={0 0 0},backref=false,colorlinks=true]
 {hyperref}
\hypersetup{linkcolor=[rgb]{0.1,0.4,0.1}, linktoc=page}
\hypersetup{citecolor=[rgb]{0.6,0,0.1}}
\hypersetup{urlcolor=[rgb]{0.1,0,0.6}}

\DeclareTextFontCommand{\emph}{\sl}

\usepackage{setspace}
\usepackage[hang]{footmisc}
\usepackage{lipsum}

\makeatletter
\renewcommand{\overset}[3][0ex]{%
  \mathrel{\mathop{#3}\limits^{
    \vbox to#1{\kern-2\ex@
    \hbox{$\scriptstyle#2$}\vss}}}}
\makeatother

\definecolor{capri}{rgb}{0.0, 0.75, 1.0}

\definecolor{ggreen}{rgb}{0.05,0.45,0.1}

\AtBeginDocument{%
    \newwrite\bibnotes
    \def\bibnotesext{Notes.bib}
    \immediate\openout\bibnotes=\jobname\bibnotesext
    \immediate\write\bibnotes{@CONTROL{REVTEX41Control}}
    \immediate\write\bibnotes{@CONTROL{%
    apsrev41Control,author="08",editor="1",pages="1",title="0",year="1"}}
     \if@filesw
     \immediate\write\@auxout{\string\citation{apsrev41Control}}%
    \fi
}%

\renewcommand{\emph}[1]{\textit{#1}}

\begin{document}

\title{Energy of cosmological spacetimes and perturbations:\\ 
a quasilocal approach}

\thanks{\setstretch{0.9}Sec. \ref{sec:qf} and \ref{sec:flrw} of this paper are an expanded version of the essay ``Quasilocal conservation laws in cosmology: a first look'' \cite{oltean_quasilocal_2020}, awarded Honorable Mention in the Gravity Research Foundation 2020 Awards for Essays on Gravitation.}

\author{Marius Oltean}
\email[]{oltean@ice.cat}
\thanks{corresponding author.}

\affiliation{\mbox{Department of Operations, Innovation and Data Sciences,}\\\vspace{-0.075cm}
\mbox{ESADE Business School, Av. Torreblanca 59, 08172 Sant Cugat (Barcelona), Spain}}

\affiliation{\mbox{Institute of Space Studies of Catalonia (IEEC),}\\\vspace{-0.075cm}
\mbox{Carrer del Gran Capit\`{a}, 2-4, Edifici Nexus, despatx 201, 08034 Barcelona, Spain}}

\author{Hossein Bazrafshan Moghaddam}
\email[]{hbazrafshan@um.ac.ir}

\affiliation{\mbox{Department of Physics, Faculty of Science, Ferdowsi University of Mashhad, Mashhad, Iran}}

\author{Richard J. Epp}
\email[]{rjepp@uwaterloo.ca}

\affiliation{\mbox{Department of Physics and Astronomy, University of Waterloo,}\\\vspace{-0.075cm}
\mbox{200 University Avenue West, Waterloo, Ontario N2L 3G1, Canada}}

\date{\today}

\begin{abstract}

\begin{spacing}{1}

Quasilocal definitions of stress-energy-momentum---that is, in the form of boundary densities (rather than local volume densities)---have proven generally very useful in formulating and applying conservation laws in general relativity. In this paper, we present a detailed application of such definitions to cosmology, specifically using the Brown-York quasilocal stress-energy-momentum tensor for matter and gravity combined. We compute this tensor, focusing on the energy and its associated conservation law, for FLRW spacetimes with no pertubrations and with scalar cosmological perturbations. For unperturbed FLRW spacetimes, we emphasize the importance of the vacuum energy (for both flat and curved space), which is almost universally underappreciated (and usually ``subtracted''), and discuss the quasilocal interpretation of the cosmological constant. For the perturbed FLRW spacetime, we show how our results recover or relate to the more typical effective local treatment of energy in cosmology, with a view towards better studying the issues of the cosmological constant and of cosmological back-reactions.

\end{spacing}

\end{abstract}

\maketitle

\pagebreak







\section{Introduction}

Observations of the Cosmic Microwave Background (CMB) and galaxy surveys
support the isotropy and homogeneity of our Universe on large scales
($>100\,{\rm {Mpc}}$) \cite{Akrami:2018vks}. Hence the maximally
symmetric class of solutions of Friedman-Lemaître-Robertson-Walker
(FLRW) for the Einstein equation is a good fit for describing our
Universe on those scales. The same observations also support the fact
that the particular FLRW solution describing our Universe is dominated
today by an exponential expansion of space. This phenomenon, which
continues to provoke a variety of theoretical problems and diverse
explanations, is most simply accounted for by the inclusion of a positive
cosmological constant $\Lambda$ in the Einstein equation\footnote{\setstretch{0.9}We work in the $(-+++)$ signature of spacetime,
in geometrized units ($G=1=c$), and follow the conventions of Wald
\cite{wald_general_1984}. In particular, Latin letters are used for
abstract spacetime indices ($a,b,c...=0,1,2,3$). 
}, 
\begin{equation}
G_{ab}[g_{cd}]+\Lambda g_{ab}=8\pi T_{ab}[g_{cd},\varphi]\,,\label{eq:EE}
\end{equation}
where $g_{ab}$ is the spacetime metric, $G_{ab}$ the Einstein tensor of this metric 
and $T_{ab}$ the matter stress-energy-momentum tensor of a collection
of matter fields $\varphi$. The typical physical interpretation given
to the cosmological constant term follows by moving it to the RHS:
\begin{equation}
G_{ab}[g_{cd}]=8\pi(T_{ab}[g_{cd},\varphi]+T_{ab}^{\Lambda}[g_{cd}])\,,
\end{equation}
where we have defined 
\begin{equation}
T_{ab}^{\Lambda}[g_{cd}]:=-\frac{1}{8\pi}\Lambda g_{ab}\,,
\end{equation}
thus interpreted from this perspective as playing the role of an effective
local stress-energy-momentum of the ``gravitational vacuum'', with
a constant local energy volume density $\rho_{\Lambda}=T_{00}^{\Lambda}$,
and equal but negative local pressure $p_{\Lambda}=-\rho_{\Lambda}$.

However it has long been understood that, fundamentally, gravitational
energy-momentum cannot be treated as a local concept in general relativity.
The difficulties that this generally implicates were long recognized by Einstein
both during and after the development of the theory, and Noether proposed
her famous conservation theorems strongly motivated by and in support
of this very claim; see e.g. Ref. \cite{chen_gravitational_2015} for more historical background. Alongside
the mathematical arguments, there is a simple physical explanation
for the non-localizability of gravitational energy-momentum, which
comes directly from the equivalence principle (see, e.g., Sec.
20.4 of Ref. \cite{misner_gravitation_1973}): if in any given locality
one is free to transform to a frame of reference with a vanishing
local ``gravitational field'' (connection coefficients), then any
local definition of (changes in) the energy-momentum of that field
would likewise have to vanish (even in situations where such changes
are physically expected).

Thus the interpretation of a $\Lambda$ term in the Einstein equation
as describing a local ``gravitational energy'', or local ``energy
of the gravitational vacuum'' can be made sense of at best only as
an effective one. Fundamentally, such notions cannot be local in character,
and the general solution taken by relativists today, though no consensus
exists on its exact formulation, is to treat them quasilocally: as
boundary rather than volume densities. Quasilocal energy-momentum notions in cosmology have not been extensively
developed up to the present work, and thus it is our aim in this paper
to offer an exploration of the usefulness of these ideas, specifically employing
the Brown-York quasilocal stress-energy-momentum tensor (for matter
and gravity) \cite{brown_quasilocal_1993} and a construction called quasilocal frames, first introduced in Ref. \cite{epp_rigid_2009} and subsequently developed in Refs. \cite{epp_existence_2012,mcgrath_quasilocal_2012,epp_momentum_2013,mcgrath_post-newtonian_2014,mcgrath_rigid_2014,oltean_geoids_2016,oltean_motion_2019,oltean_study_2019}.

Also connected to the gravitational energy-momentum issue in cosmology is the problem of the back-reaction of cosmological perturbations. At the
same time that we observe a $\Lambda$-dominated homogeneous and isotropic
(FLRW) Universe on large scales, we also observe large inhomogeneities
on small scales (such as voids and non-uniform distributions of galaxies
and stars). Then the reader with an ``average'' background in cosmology may wonder: how do these
local inhomogeneities add up to make a homogeneous Universe on large
scales? The standard answer that cosmologists provide is to say: by
``averaging''! Although there does not exist a consensus in the
field on the method and formalism by which to do this, and disputes
continue on the issue (see, e.g., Refs. \cite{Buchert:2017obp,Buchert:2015noproof,Green-Wald:2014aga,abramo_energy-momentum_1997,Paranjape:2009zu}),
there is nevertheless a common understanding that the deviations from
FLRW on small scales can get ``averaged out'' to provide a homogeneous
and isotropic Universe on large scales. Due to the nonlinear nature
of the Einstein equation, these perturbations may back-react upon the background,
requiring a careful consideration of this issue.

Consider again the Einstein equation (\ref{eq:EE}) for a perturbed
metric $g_{ab}=g_{ab}^{(0)}+\lambda g_{ab}^{(1)}+\mathcal{O}(\lambda^{2})$,
with $g_{ab}^{(0)}$ being the FLRW metric (the perturbative background)
and $\lambda$ denoting the formal (``small'') perturbation parameter.
The standard approach \cite{Green-Wald:2014aga,abramo_energy-momentum_1997}
to describing the back-reaction of the metric perturbation $g_{ab}^{(1)}$
upon $g_{ab}^{(0)}$ is to expand this equation to second order in
$\lambda$ and to take the spatial average of both sides, with the
assumption that all perturbations (the linear as well as the quadratic
metric and matter perturbations) average to zero over all of three-space.
One thus obtains:
\begin{align}
 & (G_{ab}^{(0)}[g_{cd}^{(0)}]+\Lambda g_{ab}^{(0)})+\lambda^{2}\langle G_{ab}^{(2)}[g_{cd}^{(1)}]\rangle_{\Sigma_{(0)}}+\mathcal{O}(\lambda^{3})\\
 & =8\pi(T_{ab}^{(0)}[g_{cd}^{(0)},\varphi^{(0)}]+\lambda^{2}\langle T_{ab}^{(2)}[g_{cd}^{(1)},\varphi^{(1)}]\rangle_{\Sigma_{(0)}})+\mathcal{O}(\lambda^{3})\,,
\end{align}
where $\langle\cdot\rangle_{\Sigma_{(0)}}$ indicates spatial averaging
over a background Cauchy surface $\Sigma_{(0)}$. If we now neglect
the cubic perturbation terms and set the formal perturbative parameter
$\lambda$ to $1$, the above equation can be rearranged as
\begin{equation}
G_{ab}^{(0)}[g_{cd}^{(0)}]+\Lambda g_{ab}^{(0)}=8\pi(T_{ab}^{(0)}[\varphi^{(0)}]+t_{ab}[g_{cd}^{(1)},\varphi^{(1)}])\,,
\end{equation}
in other words, the Einstein equation for the background metric $g_{ab}^{(0)}$
with an effective local (quadratic) perturbative correction to the
background local matter stress-energy-momentum tensor given by\footnote{\setstretch{0.9}This follows the notation of Ref. \cite{Green-Wald:2014aga};
in Ref. \cite{abramo_energy-momentum_1997} instead this object is
denoted as ``$\tau_{ab}$'', however we will reserve this notation
for another object in this paper.} 
\begin{equation}
t_{ab}[g_{cd}^{(1)},\varphi^{(1)}]:=\langle-\tfrac{1}{8\pi}G_{ab}^{(2)}[g_{cd}^{(1)}]+T_{ab}^{(2)}[g_{cd}^{(1)},\varphi^{(1)}]\rangle_{\Sigma_{(0)}}\,,\label{eq:t_ab}
\end{equation}
which may in this way be viewed as describing both gravitational and
non-gravitational (i.e. matter) perturbative back-reactions. 

There are different approaches\footnote{\setstretch{0.9}See also Ref. \cite{petrov_conserved_2002} for an approach to conservation laws for cosmological perturbations using the notion of ``superpotentials''.} to study these back-reactions with
different and even opposite conclusions\footnote{\setstretch{0.9}The two main camps continue to disagree on whether
the back-reactions of inhomogeneities can contribute significantly
or not to the evolution of the background Universe, and thus for example
whether or not these can act like dark energy or dark matter on different
scales. See the series of correspondence in \cite{Green-Wald:2014aga,Buchert:2015noproof,Nambu:2000qv,Green:2013yua,Brunswic:2020bjx,Buchert:2019mvq,Vigneron:2019dpj,Heinesen:2018vjp,Buchert:2017obp,Green:2016cwo,Green:2015bma}} on their significance for the evolution of the background (e.g.,
see Refs. \cite{abramo_energy-momentum_1997,Green-Wald:2014aga,Buchert:2015noproof,Nambu:2000qv}).
The disputes are either on approximation methods or the issue of the
definition of gauge-dependent variables. There is also the general
issue of locality: a space-averaged quantity is not local and a local
observer would not distinguish it \cite{Unruh:1998ic,Geshnizjani:2002wp}.

To our knowledge, the first work to consider quasilocal energy-momentum definitions applied to cosmology was Ref.~\cite{chen_quasi-local_2007}, which presented a covariant Hamiltonian approach to quasilocal notions in general, and specifically computed the quasilocal energy of unperturbed FLRW spacetimes for co-moving observers. (See also Ref. \cite{nester_energy_2008}.) Then, a consideration of the ``dark energy'' problem from a quasilocal point of view was put forward in Ref. \cite{wiltshire_gravitational_2008}. (See also Ref. \cite{wiltshire_what_2011}.) The co-moving quasilocal energy of unperturbed FLRW spacetimes was computed again and discussed at greater length in 
Ref.~\cite{afshar_quasilocal_2009}, using the quasilocal energy definition of Brown-York~\cite{brown_quasilocal_1993} as well as that of 
Epp~\cite{epp_angular_2000}. At the end of Sec.~\ref{sec:flrw} we argue that the deficiency of the Brown-York energy, as claimed in Ref.~\cite{afshar_quasilocal_2009}, is actually not a deficiency, but a masquerading of matter energy as curved space vacuum energy.

More recently, Refs. \cite{faraoni_newtonian_2015,faraoni_turnaround_2015,faraoni_beyond_2017} applied the Hawking-Hayward definition of quasilocal energy \cite{hawking_gravitational_1968,hayward_quasilocal_1994} to study three problems in cosmology: respectively, Newtonian simulations of large scale structure formation, the turnaround radius in the present accelerating universe, and lensing by the cosmological constant. Then Ref. \cite{lapierre-leonard_cosmological_2017} considered the same three cosmological problems but using the Brown-York quasilocal energy instead. The authors found that their results for these problems are unaffected by the quasilocal energy definition choice (Hawking-Hayward or Brown-York). However these involve working in perturbation theory strictly at linear order. In this paper, we will compute the energy to second order in the linear cosmological perturbations.

Concurrently with the appearance of the present work, Ref. \cite{combi_relativistic_2020} also appeared focusing on rigidity theorems in cosmology and using the notion of rigid quasilocal frames~\cite{epp_rigid_2009}. A computation of the unperturbed FLRW quasilocal energy using the Brown-York definition is presented there as well, albeit once again with the interpretation of the quasilocal vacuum energy as a ``subtraction term,'' which we argue against in Sec.~\ref{sec:flrw}.

In this paper, by employing the Brown-York quasilocal (matter plus gravitational) stress-energy-momentum tensor along with the notion of quasilocal frames, we calculate the total (matter plus gravitational) energy of cosmological spacetimes. We do this for unperturbed FLRW spacetimes with the issue of the cosmological constant in view, and for the scalar modes of cosmological perturbations with the issue of cosmological back-reactions in view. Our approach is exact and geometrical, and we make a connection to known effective local results by series expanding our quasilocal results in a ``small locality'' (for spacetime regions of small areal radius).

This paper is organized as follows. In Sec. \ref{sec:qf}, we present an overview of the gravitational energy-momentum issue in general relativity as well as the quasilocal approach to it, in self-contained technical detail for our purposes in this work. Then we compute and discuss the quasilocal energy of unperturbed FLRW spacetimes with a cosmological constant in Sec. \ref{sec:flrw}, and of scalar cosmological perturbations in Sec. \ref{sec:pert}. Finally in Sec. \ref{sec:concl} we offer some concluding remarks and outlook to future work.

\subsection*{Notation and Conventions\label{ssec:notation}}

We work in the $(-,+,+,+)$ signature of spacetime. Script upper-case
letters ($\mathscr{A}$, $\mathscr{B}$, $\mathscr{C}$, ...) are
reserved for denoting mathematical spaces (manifolds, curves, etc.).
The $n$-dimensional Euclidean space is denoted as usual by $\mathbb{R}^{n}$,
the $n$-sphere of radius $r$ by $\mathbb{S}_{r}^{n}$, and the unit
$n$-sphere by $\mathbb{S}^{n}=\mathbb{S}_{1}^{n}$. For any two spaces
$\mathscr{A}$ and $\mathscr{B}$ that are topologically equivalent
(i.e. homeomorphic), we indicate this by writing $\mathscr{A}\simeq\mathscr{B}$.

We follow the conventions of Ref.~\cite{wald_general_1984}, such that any $(k,l)$-tensor in any $(3+1)$-dimensional
(Lorentzian) spacetime $\mathscr{M}$ is 
denoted 
using the abstract index notation $A^{a_{1}\cdots a_{k}}\,_{b_{1}\cdots b_{l}}$
,
with Latin letters from the beginning of the alphabet ($a$, $b$,
$c$, ...) being used for the abstract spacetime indices ($0,1,2,3$). The components of this tensor
in a particular choice of coordinates $\{x^{\alpha}\}_{\alpha=0}^{3}$
are denoted by $A^{\alpha_{1}\cdots\alpha_{k}}\,_{\beta_{1}\cdots\beta_{l}}$, that is,
using Greek (rather than Latin) letters from the beginning of the
alphabet ($\alpha$, $\beta$, $\gamma$, ...). Spatial indices on
an appropriately defined (three-dimensional Riemannian spacelike)
constant time slice of $\mathscr{M}$ are denoted using Latin letters
from the middle third of the alphabet in Roman font: in lower-case
($i$, $j$, $k$, ...) if they are abstract, and in upper-case ($I$,
$J$, $K$, ...) if a particular choice of coordinates $\{x^{I}\}_{I=1}^{3}$
has been made.

For any $n$-dimensional manifold $\mathscr{U}$ with metric determinant $g$,
we denote its natural volume form by
\begin{equation}
\bm{\epsilon}_{\mathscr{U}}^{\,}=\sqrt{\left|g\right|}\;{\rm d}x^{1}\wedge\cdots\wedge{\rm d}x^{n}\,.\label{eq:vol_form}
\end{equation}

Let $\mathscr{S}\simeq\mathbb{S}^{2}$ be any
(Riemannian) closed two-surface that is topologically a two-sphere. Latin letters from the middle
third of the alphabet in Fraktur font ($\mathfrak{i}$, $\mathfrak{j}$,
$\mathfrak{k}$, ...) are reserved for indices of tensors on $\mathscr{S}$.
For erxample, $\epsilon_{\mathfrak{ij}}^{\mathbb{S}^{2}}$ is the volume form of the unit two-sphere $\mathbb{S}^{2}$; in standard spherical coordinates $\{\theta,\phi\},$ this
is simply given by 
\begin{equation}
\bm{\epsilon}_{\mathbb{S}^{2}}^ {}=\sin\theta\,{\rm d}\theta\wedge{\rm d}\phi\,.\label{eq:S2_volume_form}
\end{equation}
\section{Setup: Quasilocal frames and conservation laws}\label{sec:qf}

\subsection{Background and motivation}

The problem of defining gravitational energy-momentum is an old and
subtle one, the precise resolution of which still lacks a general
consensus among relativists today \cite{szabados_quasi-local_2004,jaramillo_mass_2011}\footnote{\setstretch{0.9}The author of the review \cite{szabados_quasi-local_2004} summarizes the
status of this issue: ``Although there are several promising and
useful suggestions, we not only have no ultimate, generally accepted
expression for the energy-momentum ... but there is not even a consensus
in the relativity community on general questions ... or on the list
of the criteria of reasonableness of such expressions.”}. Nevertheless, it is widely accepted that in the spatial infinity
limit of an asymptotically-flat vacuum spacetime, any proposals for
such definitions should recover the ADM definitions \cite{arnowitt_dynamics_1962,regge_role_1974}.
For example, the well-known ADM energy $\mathtt{E}_{\textrm{ADM}}$
of an asymptotically-flat vacuum spacetime is given by the integral
over a closed two-surface $\mathscr{S}\simeq\mathbb{S}_{r}^{2}$ (topologically
a two-sphere $\mathbb{S}_{r}^{2}$ of areal radius $r$) at spatial
infinity ($r\rightarrow\infty$) of an \textit{energy surface density}
(energy per unit area), given up to a factor by the trace $k$ of
the extrinsic curvature of that surface\footnote{\setstretch{0.9}In fact, taking the $r\rightarrow\infty$ limit of the integral on the RHS causes it to diverge, and so to remedy this, a common practice is to subtract from $k$ a ``reference'' boundary energy surface density, specifically the boundary extrinsic curvature of Minkowski space. We comment further on this in Subsec. \ref{subsec:cons_laws} and footnote 6.}, 
\begin{equation}
\mathtt{E}_{\textrm{ADM}}=-\frac{1}{8\pi}\lim_{r\rightarrow\infty}\oint_{\mathscr{S}}\bm{\epsilon}_{\mathscr{S}}^{\,}\,k\,.\label{eq:E_ADM}
\end{equation}

At the most basic level, such an interpretation can be conferred upon
the term above by virtue of its being the value of the Hamiltonian
of general relativity evaluated for solutions of the theory (i.e.
satisfying the canonical constraints). Similar Hamiltonian arguments
can be used to also define a general ADM four-momentum.

The ADM definitions have proven widely useful in practice, but in
principle are limited to determining the gravitational energy-momentum
of an \textit{entire} (asymptotically-flat vacuum) spacetime. Various
current proposals exist \cite{szabados_quasi-local_2004,jaramillo_mass_2011}
for the gravitational energy-momentum of \textit{arbitrary} spacetime
regions within arbitrary spacetimes. These have generally retained
the basic mathematical form (with an exact recovery in the appropriate
limit) of the ADM definitions, i.e. that of closed two-surface integrals
of \textit{surface densities} (in lieu of three-volume integrals of
volume densities, as in pre-relativistic physics), and are for this
reason referred to as \textit{quasilocal} (in lieu of local) definitions.

\begin{figure}
\begin{centering}
\includegraphics[scale=0.9]{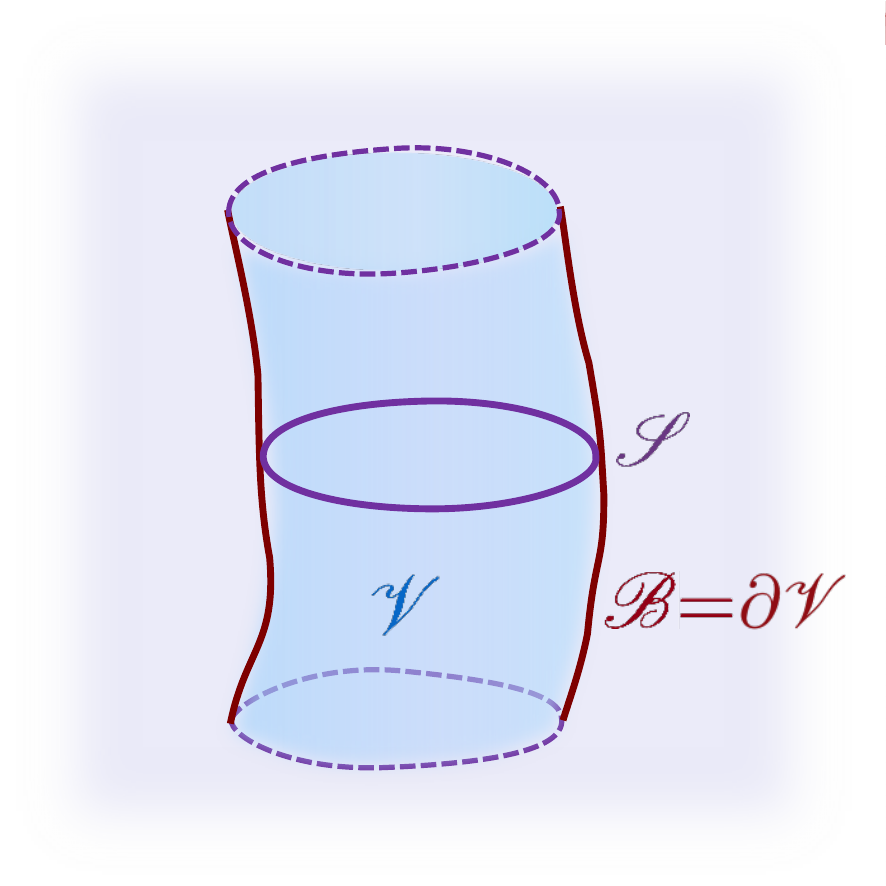} 
\par\end{centering}
\caption{A $(2+1)$ picture of a worldtube $\mathscr{V}$ with boundary $\mathscr{B}=\partial\mathscr{V}$.
A spatial slice of the latter is denoted by $\mathscr{S}$ (a closed
two-surface, topologically a two-sphere).}
\label{fig-qc-V} 
\end{figure}

In this paper, we assume and work with the quasilocal stress-energy-momentum
tensor proposed by Brown and York \cite{brown_quasilocal_1993}. While
these authors initially proposed it on the basis of a Hamilton-Jacobi
analysis, its definition can more simply be motivated by the following
argument which initially appeared in Ref. \cite{epp_momentum_2013}.
Recall that the stress-energy-momentum tensor of matter \textit{alone}
is defined from the matter action $S_{\textrm{matter}}$, up to a
factor, as $T_{ab}\propto\delta S_{\textrm{matter}}/\delta g^{ab}$,
which is a local tensor (living in the bulk). Following a similar
logic, consider a \textit{total} (matter plus gravitational) action,
\begin{equation}
S_{\textrm{total}}=S_{\textrm{matter}}+S_{\textrm{gravity}}\,.\label{eq:S_total}
\end{equation}
The gravitational action $S_{\textrm{gravity}}$ is, in any spacetime
region $\mathscr{V}$—which for simplicity henceforth we take to be
a worldtube\footnote{\setstretch{0.9}This assumption is made here only to simplify our
motivating discussion. We use the word ``worldtube'' to refer to
a four-dimensional spacetime region with topology $\mathbb{R}\times\mathbb{B}^{3}$
where $\mathbb{B}^{3}$ is the three-ball. For a full analysis for
a completely arbitrary $\mathscr{V}$, see e.g. \cite{brown_action_2002}.}, i.e. the history of a finite spatial three-volume, see Fig. \ref{fig-qc-V}—as
a sum, 
\begin{equation}
S_{\textrm{gravity}}=S_{\textrm{EH}}+S_{\textrm{GHY}}\,.\label{eq:S_gravity}
\end{equation}
The first term is the Einstein-Hilbert (bulk) term, 
\begin{equation}
S_{\textrm{EH}}=\frac{1}{16\pi}\int_{\mathscr{V}}\bm{\epsilon}_{\mathscr{V}}^{\,}\,R\,,\label{eq:S_EH}
\end{equation}
and the second is the Gibbons-Hawking-York (boundary) term, 
\begin{equation}
S_{\textrm{GHY}}=-\frac{1}{8\pi}\int_{\mathscr{B}}\bm{\epsilon}_{\mathscr{B}}^{\,}\,K\,,\label{eq:S_GHY}
\end{equation}
where $K$ is the trace of the extrinsic curvature of the boundary
$\mathscr{B}=\partial\mathscr{V}\simeq\mathbb{R}\times\mathbb{S}^{2}$.
Now from the total action (\ref{eq:S_total}), a \textit{total} stress-energy-momentum
tensor $\tau_{ab}$ can be defined (analogously to the matter-only
$T_{ab}$), as $\tau_{ab}\propto\delta S_{\textrm{total}}/\delta g^{ab}$.
Assuming the Einstein equation (\ref{eq:EE}) holds in $\mathscr{V}$,
the bulk term in the functional derivative now vanishes, and the result
evaluates to a tensor \textit{living on the boundary} $\mathscr{B}$,
i.e. a quasilocal tensor, known as the \textit{Brown-York tensor},
and given (with the appropriate proportionality factor restored) by
\begin{equation}
\tau_{ab}=-\frac{1}{8\pi}\Pi_{ab}\,,\label{eq:tau}
\end{equation}
where $\Pi_{ab}$ is the canonical momentum (defined in the usual
way from the extrinsic curvature) of $\mathscr{B}$. This expresses
boundary densities of the total (matter plus gravitational) energy-momentum
which, when integrated over the two-surface intersection of a spacelike
Cauchy slice and $\mathscr{B}$, yield the total values thereof contained
in the part of the Cauchy slice (three-volume) within $\mathscr{B}$.

\subsection{Quasilocal frames}\label{sec:QF}

Conservation laws for energy, momentum and angular momentum using
the Brown-York tensor have been formulated with the use of a concept
called \textit{quasilocal frames} \cite{epp_rigid_2009,epp_existence_2012,mcgrath_quasilocal_2012,epp_momentum_2013}, which has been subsequently applied to post-Newtonian theory \cite{mcgrath_post-newtonian_2014}, relativistic geodesy \cite{oltean_geoids_2016} and the gravitational self-force problem \cite{oltean_motion_2019,oltean_study_2019}.

\begin{figure}
\begin{centering}
\includegraphics[scale=0.9]{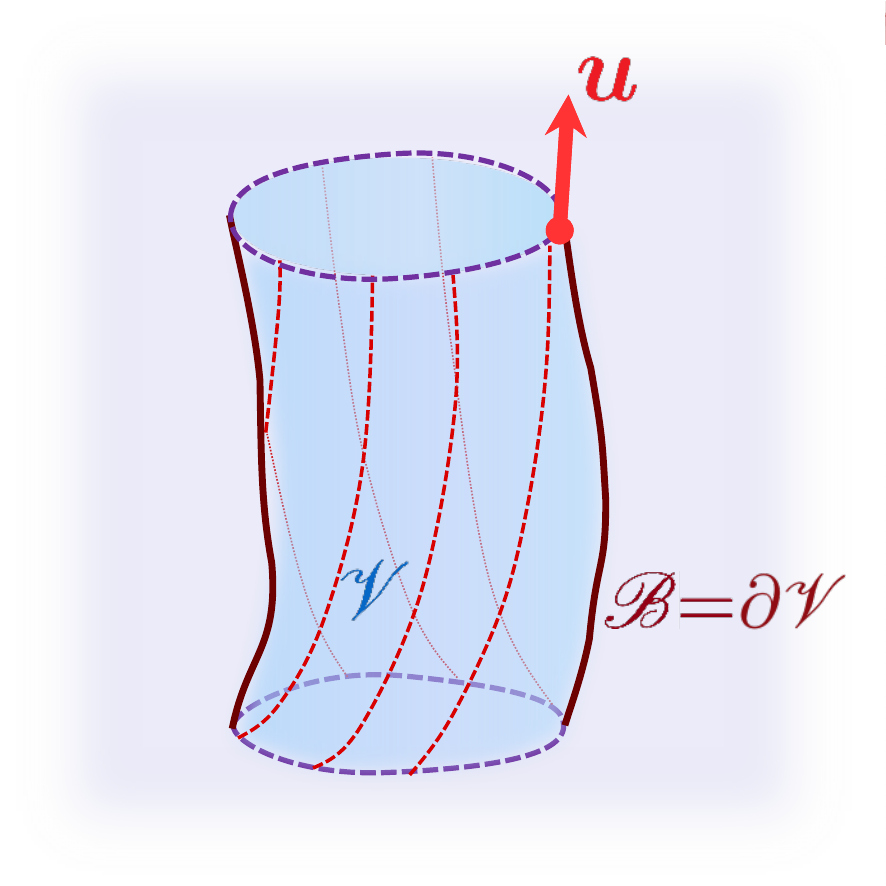} 
\par\end{centering}
\caption{A $(2+1)$ picture of a quasilocal frame $(\mathscr{B},u^{a})$, where
$\mathscr{B}=\partial\mathscr{V}$ is the boundary of a worldtube
$\mathscr{V}$ and $u^{a}$ is the timelike four-velocity of a two-parameter
family of observers the integral curves of which constitute $\mathscr{B}$.}
\label{fig-qc-Vqf} 
\end{figure}

Essentially, the idea is that additional structure is required on
$\mathscr{B}$ in order to specify the components of stress-energy-momentum
seen by a particular set of observers on $\mathscr{B}$. In particular,
what is required is a two-parameter congruence with timelike observer
four-velocity $u^{a}\in T\mathscr{B}$, the integral curves of which
constitute $\mathscr{B}$. Such a pair $(\mathscr{B},u^{a})$ is referred
to as a quasilocal frame. See Fig. \ref{fig-qc-Vqf}. 

With this in hand, the Brown-York tensor $\tau_{ab}$ can now be decomposed
(analogously to the local matter stress-energy-momentum tensor $T_{ab}$)
into components representing the quasilocal energy, momentum and stress
respectively: 
\begin{align}
\mathcal{E}=\, & u^{a}u^{b}\tau_{ab}\,,\\
\mathcal{P}^{a}=\, & -\sigma^{ab}u^{c}\tau_{bc}\,,\\
\mathcal{S}^{ab}=\, & -\sigma^{ac}\sigma^{bd}\tau_{cd}\,.
\end{align}
(Equivalently, $\tau^{ab}=u^{a}u^{b}\mathcal{E}+2u^{(a}\mathcal{P}^{b)}-\mathcal{S}^{ab}$.)
Here, $\sigma_{ab}$ is the metric induced on the tangent space of
$\mathscr{B}$ orthogonal to $u^{a}$. In particular,
\begin{equation}
\sigma_{ab}=g_{ab}-n_{a}n_{b}+u_{a}u_{b}\,,
\end{equation}
where $n^{a}$ is the unit normal to $\mathscr{B}$. We thus have,
e.g., the following general expression for the quasilocal energy surface
density: 
\begin{equation}
\mathcal{E}=\tau_{ab}u^{a}u^{b}=-\frac{1}{8\pi}k\,,\label{eq:Egeneral}
\end{equation}
where $k$ is the observers' (two-dimensional) spatial trace of the
extrinsic curvature of $\mathscr{B}$. (Note that this is readily
reminiscent of the ADM definitions, but in principle applicable here
to any spacetime region.)

Before we proceed, we add that in general the four-velocity of our
congruence $u^{a}$ need not be chosen to be orthogonal to the constant
time slices $\mathscr{S}$ foliating $\mathscr{B}$. Denoting by $\tilde{u}^{a}$
the timelike unit normal to $\mathscr{S}$, we will in general have
a shift between these:
\begin{equation}
\tilde{u}^{a}=\gamma(u^{a}+v^{a})\,,\label{eq:u^tilde}
\end{equation}
where $v^{a}$ represents the spatial two-velocity of fiducial observers
that are at rest with respect to $\mathscr{S}$ as measured by our
congruence of quasilocal observers, and $\gamma=1/\sqrt{1-v^{a}v_{a}}$
is the Lorentz factor.

\subsection{Conservation laws}\label{subsec:cons_laws}

The starting point for constructing conservation laws from the Brown-York
tensor using quasilocal frames is the following identity (which is
simply the Leibnitz rule):
\begin{equation}
\mathcal{D}_{a}(\tau{}^{ab}\psi_{b})=(\mathcal{D}_{a}\tau^{ab})\psi_{b}+\tau^{ab}(\mathcal{D}_{a}\psi_{b})\,,\label{eq:Leibnitz}
\end{equation}
for an arbitrary vector field $\psi^{a}$ in the tangent space of
$\mathscr{B}$, with $\mathcal{D}_{a}$ representing the derivative
operator induced on $\mathscr{B}$. Integrating this equation on both
sides over a portion of $\mathscr{B}$ (between an initial and final
time) and using Stokes' theorem on the LHS produces conservation laws
for energy, momentum and angular momentum depending upon the choice
of the vector $\psi^{a}$. 

While the (linear and angular) momentum conservation laws require
a more detailed analysis \cite{epp_momentum_2013}, the one for energy can simply
be seen to arise by choosing $\psi^{a}=u^{a}$. In that case, the
LHS of the integrated equation (\ref{eq:Leibnitz}) expresses the difference between
the total energies at two different times, with the general expression
of the total energy at any given time (on any time slice $\mathscr{S}$
of $\mathscr{B}$) given by
\begin{equation}
\mathtt{E}=\int_{\mathscr{S}}\bm{\epsilon}_{\mathscr{S}}^{\,}\,\left(\mathcal{E}-\mathcal{P}^{a}v_{a}\right)\,.\label{eq:Etotgeneral}
\end{equation}

The advantage of this construction is that it permits the computation
of (changes in) these various quantities for any spacetime region
$\mathscr{V}$ on the boundary of which such a congruence (quasilocal
frame) can be defined. For any small spatial region, that is, one
contained inside a topological two-sphere having an areal radius $r$
much smaller than the spacetime curvature and scale of matter density
variation, the quasilocal energy density (\ref{eq:Egeneral}) evaluates
in general to \cite{epp_momentum_2013}: 
\begin{equation}
\mathcal{E}=\mathcal{E}_{\textrm{vac}}(r)+\mathcal{O}\left(r\right)\,,\label{eq:Esmallr}
\end{equation}
where 
\begin{equation}
\mathcal{E}_{\textrm{vac}}(r)=-\frac{1}{4\pi r}\label{eq:Evac}
\end{equation}
is known as the \textit{vacuum} quasilocal energy density. Matter
contributions to $\mathcal{E}$ begin possibly from $\mathcal{O}(r)$
and gravitational contributions possibly from $\mathcal{O}(r^{3})$.
This vacuum energy is a geometrical term, simply accountable from
the fact that the extrinsic curvature trace of a round two-sphere
in flat space is $k=2/r$, and it is often regarded in the literature
as unphysical\footnote{\setstretch{0.9}Indeed, this often relates to the argument (see
e.g. \cite{poisson_relativists_2007}) that one should subtract a
``reference'' (vacuum) action $S_{\textrm{vac}}$ from the gravitational
action $S_{\textrm{gravity}}$ evaluated over all spacetime, in particular
taking $S_{\textrm{vac}}$ to be $S_{\textrm{gravity}}$ evaluated
over all of flat space (hence divergent), and therefore work with
a ``regularized'' gravitational action $S_{\textrm{gravity}}-S_{\textrm{vac}}$.
The subtraction of such an $S_{\textrm{vac}}$ is equivalent to the
subtraction of the vacuum energy term (\ref{eq:Evac}) from the quasilocal
energy density $\mathcal{E}$. However, we emphasize that such an
argument is predicated on defining $S_{\textrm{gravity}}$ over all
of spacetime, which it should not be for properly formulating the
usual action principle, and which would thus make it (unnecessarily)
divergent by construction.} \cite{szabados_quasi-local_2004,brown_action_2002}. Yet, analyses
and applications of the quasilocal conservation laws have shown how
this term is in fact needed for a proper accounting of gravitational
energy-momentum transfer \cite{epp_momentum_2013}. In particular,
it is intimately linked to and logically self-consistent with the
existence of a \textit{vacuum pressure} (similarly, the leading term
in an expansion in $r$ of the quasilocal pressure ${\rm P}$, defined
from the observers' spatial trace of $\tau_{ab}$), 
\begin{equation}
{\rm P}_{\textrm{vac}}(r)=-\frac{1}{8\pi r}\,.\label{eq:Pvac}
\end{equation}
Physically, a negative vacuum pressure acts as a positive surface
tension on the boundary $\mathscr{S}$, resulting in a ``${\rm P}_{\textrm{vac}}{\rm d}A$''
work term, which exactly accounts for the change in negative vacuum
energy. These vacuum terms have been shown to be necessary in applications,
including recently in the gravitational self-force problem \cite{oltean_motion_2019,oltean_study_2019},
where they play a key role in accounting for the perturbative correction
to the motion of a point particle due to gravitational back-reaction.


\section{Quasilocal energy conservation in FLRW spacetimes}\label{sec:flrw}

Thus far, to our knowledge, these quasilocal quantities and their conservation laws
have not been thoroughly investigated in the context of cosmology.
In what follows, we present a basic application of the quasilocal frame energy conservation law to FLRW spacetimes with a perfect-fluid matter source, and compare our results, in particular, to those of Ref.~\cite{afshar_quasilocal_2009}. 

Integrating equation~(\ref{eq:Leibnitz}) over a portion, $\Delta\mathscr{B}$, of $\mathscr{B}$, bounded by initial and final time slices $\mathscr{S}_{\rm i}$ and $\mathscr{S}_{\rm f}$, and setting $\psi^a = u^a$ (which will be orthogonal to $\mathscr{S}_{\rm i}$ and $\mathscr{S}_{\rm f}$ in the cases we will consider in this section, i.e. here we have $v^{a}=0$), we obtain the following quasilocal energy conservation law:
\begin{equation}\label{eq:intconserve}
\int_{\mathscr{S}_{\rm f}-\mathscr{S}_{\rm i}}\bm{\epsilon}_{\mathscr{S}}^{\,}\,\mathcal{E} 
= \int_{\Delta\mathscr{B}}\bm{\epsilon}_{\mathscr{B}}^{\,}\,\left[T^{ab}u_a n_b-\tau^{ab} \mathcal{D}_a u_b\right]\, .
\end{equation}
This law says that the change in the quasilocal energy between the initial and final time slices is due to a flux of matter energy (the $T^{ab}n_a u_b$ term on the RHS) plus a flux of gravitational energy (the $-\tau^{ab} \mathcal{D}_a u_b$ term on the RHS) through $\Delta\mathscr{B}$.\footnote{\setstretch{0.9}These are radially \emph{inwards} fluxes; if they are positive, they cause the quasilocal energy to increase.} We will consider two complementary examples of quasilocal frames in FLRW cosmology: \emph{co-moving} observers who reside on a round sphere of dynamic areal radius and see only a gravitational energy flux (zero matter energy flux), and \emph{rigid} observers who reside on a round sphere of constant areal radius and see only a matter energy flux (zero gravitational energy flux).

We begin by expanding the gravitational energy flux density using the decomposition $\tau^{ab}=u^{a}u^{b}\mathcal{E}+2u^{(a}\mathcal{P}^{b)}-\mathcal{S}^{ab}$ introduced in Sec.~\ref{sec:QF} above:
\begin{equation}\label{eq:expandgravflux}
-\tau^{ab} \mathcal{D}_a u_b = \mathcal{S}^{ab}\theta_{ab}-\alpha^a\mathcal{P}_a = \mathcal{S}^{ab}_{\rm TF}\,\theta_{ab}^{\rm TF} + {\rm P}\theta -\mathcal{P}^a\alpha_a \, .
\end{equation}
Here $\theta_{ab}=\sigma_{(a}^{\;\;\; c}\sigma_{b)}^{\;\;\;d}\nabla_c u_d$ is the strain rate tensor, which is split into the expansion, $\theta$ (trace part), and shear, $\theta_{ab}^{\rm TF}$ (trace-free part), of the two-parameter $u^a$ congruence; ${\rm P}$ is the quasilocal pressure; and $\alpha_a=\sigma_a^{\;\;b}a_b$ is the projection of the observers' four-acceleration $a_a=u^b\nabla_b u_a$ tangent to $\mathscr{B}$. The term $\mathcal{S}^{ab}\theta_{ab}$ is of the standard form ``stress'' $\times$ ``strain rate'' $=$ ``power per unit area'' in an elastic medium; it splits into a gravitational radiation term, $\mathcal{S}^{ab}_{\rm TF}\,\theta_{ab}^{\rm TF}$ ($\mathcal{S}^{ab}_{\rm TF}$ and $\theta_{ab}^{\rm TF}$ can begin at quadrupole order), and another gravitational energy flux term, ${\rm P}\theta$ (${\rm P}$ and $\theta$ can begin at monopole order). The term $\mathcal{P}^a\alpha_a$ represents a gravitational energy flux due to the observers accelerating relative to a quasilocal momentum density. This term has a direct analogue in both Newtonian and special relativistic mechanics (e.g., an observer accelerating towards a non-accelerating object increases the object's observed energy at a rate equal to the scalar product of the observer's acceleration and the object's observed momentum). There are five functional degrees of freedom in $\theta_{ab}^{\rm TF}$, $\theta$, and $\alpha_a$, generically three of which can be ``gauged away'' by suitable choice of quasilocal frame. For more detailed discussions, see Refs. \cite{epp_rigid_2009,epp_existence_2012,mcgrath_quasilocal_2012,epp_momentum_2013,mcgrath_post-newtonian_2014,mcgrath_rigid_2014,oltean_geoids_2016,oltean_motion_2019,oltean_study_2019}. 

In the case of co-moving or rigid quasilocal observers in FLRW cosmology, because of the spherical symmetry, $\theta_{ab}^{\rm TF}$ and $\alpha_a$ will be zero. However, there could be a non-zero monopole expansion, $\theta$ (and in general there is always at least a vacuum quasilocal pressure). Thus, the general quasilocal energy conservation law reduces, in the cases in which we will be interested, to simply:
\begin{equation}\label{eq:intconservereduced}
\int_{\mathscr{S}_{\rm f}-\mathscr{S}_{\rm i}}\bm{\epsilon}_{\mathscr{S}}^{\,}\,\mathcal{E} 
= \int_{\Delta\mathscr{B}}\bm{\epsilon}_{\mathscr{B}}^{\,}\,\left[T^{ab}u_a n_b+{\rm P}\theta\right]\, .
\end{equation}

The FLRW line element in spherical spacetime coordinates $\{T,R,\Theta,\Phi\}$ reads:
\begin{equation}
{\rm d}s^2=-{\rm d}T^{2}+a^{2}(T)\left[\frac{{\rm d}R^{2}}{1-{\sf k}R^2}+R^{2}{\rm d}\Omega^{2}\right]\,,\label{eq:FLRW}
\end{equation}
where ${\rm d}\Omega^2$ is the line element on the unit round two-sphere
with coordinates $\{\Theta,\Phi\}$, and ${\sf k}$ is the Gaussian curvature of space when $a(T)=1$.

We will begin with an analysis of a quasilocal frame of co-moving observers, for which $\mathscr{B}\equiv\mathscr{B}_{\rm C}$ is an $R=$ constant surface.\footnote{\setstretch{0.9}All quantities associated with the co-moving quasilocal frame will have the label ``C'' (subscript or superscript, as convenience dictates).} It is easy to see that:
\begin{equation}
u^a\equiv u_{\rm C}^a = \left(\frac{\partial}{\partial T}\right)^a
\end{equation}
\begin{equation}
n^a\equiv n_{\rm C}^a = \frac{\sqrt{1-{\sf k}R^2}}{a}\left(\frac{\partial}{\partial R}\right)^a
\end{equation}

The perfect fluid matter stress-energy-momentum tensor, adapted to the co-moving observers, is written:
\begin{equation}\label{eq:FLRWSEM}
T^{ab}=\rho \, u_{\rm C}^a u_{\rm C}^b + p \, (g^{ab}+u_{\rm C}^a u_{\rm C}^b)\, . 
\end{equation}
(To reduce notational clutter, we omit the label ``C'' on the mass volume density $\rho$ and pressure $p$.) It is clear that $T^{ab}u^{\rm C}_a n^{\rm C}_b=0$: the co-moving observers see no (radial) matter energy flux through their two-sphere. However, due to the expansion of their sphere, they do see a (radial) ${\rm P}\theta$ gravitational energy flux, analogous to a ${\rm P}\,{\rm d}A$ work term. 

The quantities in (\ref{eq:intconservereduced}) are found to be the following: the integration measures are
\begin{equation}
\bm{\epsilon}_{\mathscr{S}}^{\rm C}=(aR)^2\,\bm{\epsilon}_{\mathbb{S}^{2}}^ {}\;\;\;\;\;{\rm and} \;\;\;\;\; \bm{\epsilon}_{\mathscr{B}}^{\rm C}={\rm d}T\,\wedge\bm{\epsilon}_{\mathscr{S}}^{\rm C}\,,
\end{equation}
so we see that $aR$ is the (dynamic) areal radius; the expansion of the congruence is
\begin{equation}
\theta_{\rm C} = 2\,\frac{\dot{a}}{a}=2H\,,
\end{equation}
where $H$ is the Hubble parameter; the quasilocal pressure is
\begin{equation}
{\rm P}_{\rm C}=\frac{1}{2}\,{\cal E}_{\rm C}\,,
\end{equation}
which follows from the general identity $\mathcal{E}-2{\rm P}=n\cdot a/4\pi$ and the fact that the co-moving observers experience no radial proper acceleration ($n_{\rm C}^a a^{\rm C}_a=0$); and finally, the quasilocal energy density is
\begin{equation}\label{eq:ComovingEnergyDensity}
{\cal E}_{\rm C}=-\frac{1}{4\pi aR}\sqrt{1-{\sf k}R^2}=\mathcal{E}_{\textrm{vac}}(aR)\sqrt{1-{\sf k}R^2}\,.
\end{equation}
Integrating over the two-sphere we find the co-moving quasilocal energy (i.e. the total energy contained inside the co-moving three-volume):
\begin{equation}\label{eq:ComovingEnergy}
\mathtt{E}_{\rm C}=\mathtt{E}_{\textrm{vac}}(aR)\sqrt{1-{\sf k}R^2}\,.
\end{equation}

Surprisingly, the co-moving observers appear to see only the energy of empty space, i.e. the vacuum energy contained in a round sphere of (dynamic) areal radius $aR$ in flat space, ``corrected'' by the factor $\sqrt{1-{\sf k}R^2}$ to apparently make it a curved space vacuum energy. There is no (direct) reference to the matter in the spacetime ($\rho$ or $p$). On the one hand, this is consistent with the co-moving observers seeing only a gravitational energy flux, and no matter energy flux. On the other hand, the time rate of change of the co-moving quasilocal energy, $\dot{\mathtt{E}}_{\rm C}=H\mathtt{E}_{\rm C}$, \emph{does} depend on matter via $H$, which is connected to $\rho$ and $p$ through the Friedmann equations. Moreover, the presence of ${\sf k}$ in $\sqrt{1-{\sf k}R^2}$ indicates that this term may somehow represent the ``total'' energy (in the sense that the $-{\sf k}/a^2$ term in the time-time Friedmann equation $\dot{a}^2/a^2+{\sf k}/a^2-\Lambda/3=8\pi\rho/3$ represents the ``total'' cosmological energy). We will be able to shed further light on this interesting result after considering the case of rigid quasilocal observers, to which we now turn.

To construct a quasilocal frame of rigid observers, we consider the coordinate transformation 
\begin{equation}
\begin{cases}
T & =t\,,\\
R & =r/a(t)\,,\\
\Theta & =\theta\,,\\
\Phi & =\phi\,,
\end{cases}\label{eq:coords}
\end{equation}
which takes us from a $\mathscr{B}\equiv\mathscr{B}_{\rm C}=$ constant $R$ timelike hypersurface to a $\mathscr{B}\equiv\mathscr{B}_{\rm R}=$ constant $r$ timelike hypersurface.\footnote{\setstretch{0.9}Analogously to the label ``C'', all quantities associated with the rigid quasilocal frame will have the label ``R''.} Rigid observers on a constant $r$ hypersurface at time $t$ will see co-moving observers on a constant $R$ hypersurface (with $R=r/a(t)$) at time $T=t$ moving radially outwards (if $H>0$) with a relative velocity $\beta$ (given below). See Fig.~\ref{fig-qc-frw}. In other words, the pairs ($u_{\rm C}^a, n_{\rm C}^a$) and ($u_{\rm R}^a, n_{\rm R}^a$) are related by a radial boost:
\begin{align}\label{eq:boost}
\begin{split}
u_{\rm R}^a & =\gamma (u_{\rm C}^a-\beta n_{\rm C}^a)\,,\\
n_{\rm R}^a & =\gamma (n_{\rm C}^a-\beta u_{\rm C}^a)\,,
\end{split}
\end{align}
where $\gamma=1/\sqrt{1-\beta^2}$ and
\begin{equation}\label{beta}
\beta=\frac{aRH}{\sqrt{1-{\sf k}R^2}}=\frac{rH}{\sqrt{1-{\sf k}(a/r)^2}}\,,
\end{equation}
which, of course, is just proportional to the Hubble parameter. From (\ref{eq:coords}) we see that
\begin{equation}
u_{\rm R}^a\equiv\frac{1}{N}\left(\frac{\partial}{\partial t}\right)^a=\frac{1}{N}\left(\frac{\partial}{\partial T}-RH\frac{\partial}{\partial R}\right)^a\,,
\end{equation}
so $\mathscr{B}_{\rm R}$ comprises the integral curves of $\partial/\partial t$, and the lapse function for the rigid quasilocal observers is $N=1/\gamma$, just the Lorentz time dilation associated with the relative velocity $\beta$.

\begin{figure}
\begin{centering}
\includegraphics[scale=1.1]{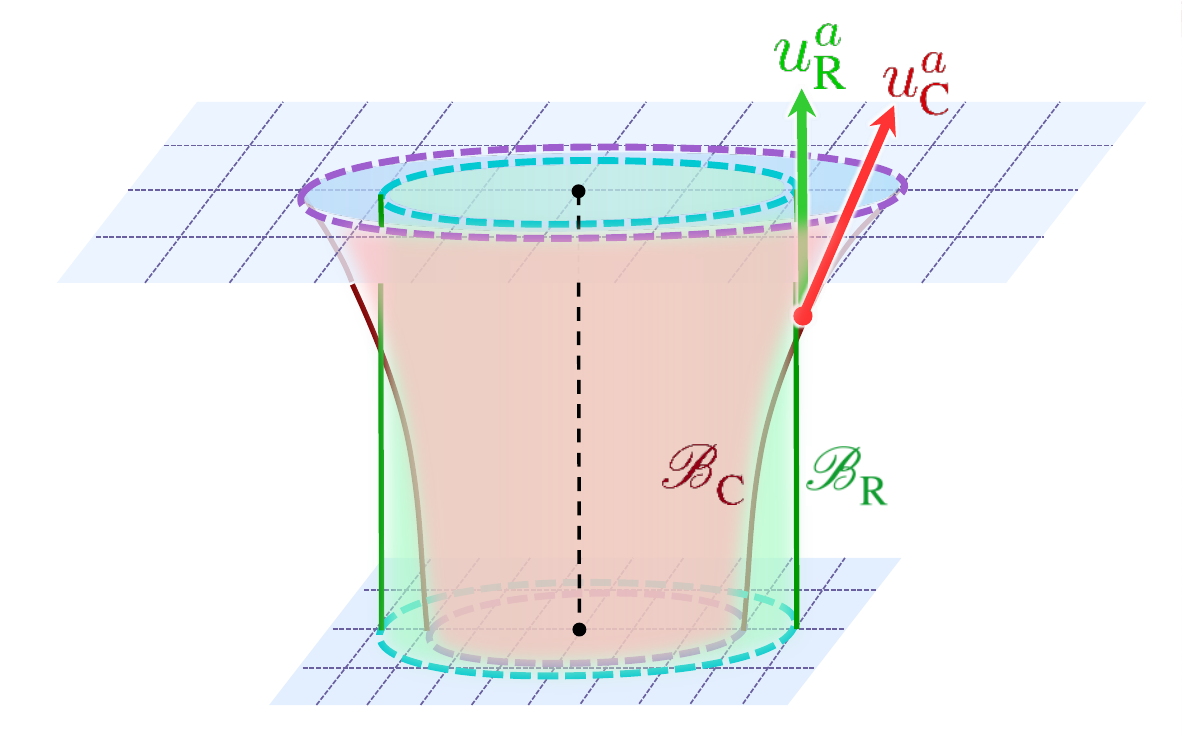} 
\par\end{centering}
\caption{A $(2+1)$ picture in a flat FLRW spacetime showing, in red, a co-moving
quasilocal frame $(\mathscr{B}_{\textrm{C}},u_{\textrm{C}}^{a})$
(with the boundary at a fixed co-moving radius) and, in green, a rigid
quasilocal frame $(\mathscr{B}_{\textrm{R}},u_{\textrm{R}}^{a})$
(with the boundary at a fixed proper radius). The black dotted line
is the center of the spatial FLRW coordinates, depicted
(as a Cartesian system) on each Cauchy slice by dotted blue lines.}
\label{fig-qc-frw} 
\end{figure}

Now we return to (\ref{eq:intconservereduced}), specializing it to the case of the rigid quasilocal frame. The integration measures are now
\begin{equation}
\bm{\epsilon}_{\mathscr{S}}^{\rm R}=r^2\,\bm{\epsilon}_{\mathbb{S}^{2}}^ {}\;\;\;\;\;{\rm and} \;\;\;\;\; \bm{\epsilon}_{\mathscr{B}}^{\rm R}=N\,{\rm d}T\,\wedge\bm{\epsilon}_{\mathscr{S}}^{\rm R}\,.
\end{equation}
As expected, the areal radius, which is $r$, is now constant, and so the expansion of the congruence is zero: $\theta_{\rm R}=0$. In contrast to the co-moving observers, the rigid observers see no gravitational energy flux. Instead, they see a \emph{matter} energy flux
\begin{equation}\label{eq:MatterEnergyFlux}
T^{ab}u^{\rm R}_a n^{\rm R}_b = -(\rho+p)\,\gamma^2\beta\,,
\end{equation}
which follows from using (\ref{eq:boost}) in (\ref{eq:FLRWSEM}).\footnote{\setstretch{0.9}There is an interesting discussion of the purely (special) relativistic physics encoded in such a matter energy flux given in Sec.~47 of~\cite{rindler_introduction_1991}.}
Note that this is a flux associated with motion relative to the ``inertial'' mass density $(\rho+p)$; it vanishes in the case of a cosmological constant perfect fluid, as it should. Finally, the quasilocal energy density is
\begin{equation}\label{eq:RigidEnergyDensity}
{\cal E}_{\rm R}=-\frac{1}{\gamma}\frac{1}{4\pi r}\sqrt{1-{\sf k}(r/a)^2}=\frac{1}{\gamma}\mathcal{E}_{\textrm{vac}}(r)\sqrt{1-{\sf k}(r/a)^2}\,.
\end{equation}
Integrating over the two-sphere we find the rigid quasilocal energy:
\begin{equation}\label{RQF_ver1}
\mathtt{E}_{\rm R}=\frac{1}{\gamma}\mathtt{E}_{\textrm{vac}}(r)\sqrt{1-{\sf k}(r/a)^2}\,.
\end{equation}
Comparing with (\ref{eq:ComovingEnergyDensity}) we see that (where $\mathscr{B}_{\rm C}$ and $\mathscr{B}_{\rm R}$ intersect)
\begin{equation}\label{eq:LorentzFactorRelation}
\mathtt{E}_{\rm C}=\gamma \,\mathtt{E}_{\rm R}\,.
\end{equation}
We interpret this as saying that $\mathtt{E}_{\rm R}$ is a sort of minimum ``rest'' mass-energy seen by ``stationary'' observers ``at rest'' in space (the rigid observers are all at rest with respect to each other, and see a static spatial two-geometry); the co-moving observers are in motion with respect to these rigid observers (and with respect to each other, and see a time-changing spatial two-geometry) and see the same ``rest'' mass-energy, but dilated by the expected Lorentz factor. What allows for such an absolute distinction between ``at rest'' and ``moving'' is the absolute (coordinate-invariant) distinction between a static spatial two-geometry and a dynamic one.

But still, we see no (direct) reference to the matter in the spacetime ($\rho$ or $p$). To see it, consider the following remarkable identity:
\begin{equation}
\frac{1}{\gamma}\sqrt{1-{\sf k}(r/a)^2}=\sqrt{1-\beta^2}\sqrt{1-{\sf k}(r/a)^2}=\sqrt{1-r^2\left(\frac{8\pi}{3}\rho_{\textrm{eff}}\right)}\,,\label{eq:sqrt_factors}
\end{equation}
which follows from (\ref{beta}) and the time-time Friedmann equation,
\begin{equation}
\frac{\dot{a}^{2}}{a^{2}}+\frac{\mathsf{k}}{a^{2}}-\frac{\Lambda}{3}=\frac{8\pi}{3}\rho\Leftrightarrow\frac{\dot{a}^{2}}{a^{2}}+\frac{\mathsf{k}}{a^{2}}=\frac{8\pi}{3}\rho_{\textrm{eff}}\,,
\end{equation}
where we have defined 
\begin{equation}
\rho_{\textrm{eff}}=\rho+\rho_{\Lambda}
\end{equation}
as the total ``effective'' (matter plus cosmological constant) local
density, with $\rho_{\Lambda}=\Lambda/8\pi$ as in the Introduction.
Notice that both square root factors in (\ref{eq:sqrt_factors}) play a crucial role. The ``kinetic'' $\beta^2$ term in the Lorentz factor gives rise to the ``kinetic'' cosmological term $\dot{a}^2/a^2$; the ``total energy'' ${\sf k}$ term in the spatial geometry factor gives rise to the ``total energy'' cosmological term $-{\sf k}/a^2$; these combine to yield the ``gravitational potential energy'' cosmological term $-8\pi\rho_{\textrm{eff}}/3$ in the final expression. Thus we have
\begin{equation}\label{eq:RQFenergy}
\mathtt{E}_{\rm R}=\mathtt{E}_{\textrm{vac}}(r)\sqrt{1-r^2
\left(\frac{8\pi}{3}\rho_{\textrm{eff}}\right)}\,.
\end{equation}

To see that this expression makes sense, consider the small $r$ expansion (in which we have temporarily, in this paragraph and the next, restored factors of $G$ and $c$):
\begin{equation}\label{eq:ERsmall_r}
\mathtt{E}_{\rm R}=-\frac{c^4}{G}\,r+\left(\frac{4\pi}{3}r^3\right)\rho_{\textrm{eff}} c^2+\left(\frac{8\pi^2}{9}r^5\right)G\rho_{\textrm{eff}}^2+\mathcal{O}(r^{7})
\end{equation}
Observe that the multiplicative flat space vacuum energy factor, $\mathtt{E}_{\textrm{vac}}(r)=-c^4 r/G$ in (\ref{eq:RQFenergy}), plays a crucial role in obtaining the correct matter energy result at $\mathcal{O}(r^{3})$: the correct power of $r$, the cancellation of the two $G$s, and the factor $c^2$. And, as expected, the gravitational energy term at $\mathcal{O}(r^{5})$ is proportional to $G\rho_{\textrm{eff}}^2 r^5$ (like the Newtonian gravitational potential energy of a ball of radius $r$ and uniform mass density, which is proportional to $GM^2/r$, or $G\rho_{\textrm{eff}}^2 r^5$). $\mathtt{E}_{\textrm{vac}}(r)$ plays a key role in non-linearly converting the ``gravitational potential energy'' square root factor in (\ref{eq:RQFenergy}) into a sensible (quasilocal) energy that includes vacuum energy, matter energy, and gravitational energy, in a simple, exact expression that manifestly reduces to pure flat space vacuum energy in the absence of matter ($\rho=0$), including a cosmological constant. Moreover, as $r$ increases, eventually $\mathtt{E}_{\rm R}$ vanishes as $\beta$ approaches one, or $r$ approaches $\sqrt{3c^2/(8\pi G\rho_{\textrm{eff}})}$. For a vacuum (no matter, $\rho=0$) FLRW Universe with only a cosmological constant, such that $\rho_{\textrm{eff}}= \rho_\Lambda=\Lambda c^2/(8\pi G)$, we have
\begin{equation}\label{eq:E_R_Lambda}
\mathtt{E}_{\rm R}^\Lambda=-\frac{c^4}{G}\,r\sqrt{1-r^2\left(\frac{\Lambda}{3}\right)}\,,
\end{equation}
which (assuming $\Lambda>0$) vanishes at both $r=0$ and the horizon $r=\sqrt{3/\Lambda}$, and is pure imaginary (undefined) for $r>\sqrt{3/\Lambda}$. All of these properties of $\mathtt{E}_{\rm R}$ are sensible, and since $\mathtt{E}_{\rm C}$ is related to $\mathtt{E}_{\rm R}$ by a sensible Lorentz factor, recall (\ref{eq:LorentzFactorRelation}), $\mathtt{E}_{\rm C}$, too, is sensible.

It is interesting to note that the small $r$ expansion of (\ref{eq:E_R_Lambda}), which is (\ref{eq:ERsmall_r}) with $\rho_{\textrm{eff}} = \rho_\Lambda$, shows that the quasilocal (matter plus gravitational) energy contained in the sphere is not merely the cosmological mass volume density, $\rho_\Lambda$, times the volume of the sphere (times $c^2$). Besides the additional (geometrical) vacuum energy contribution at order $r$, there is a \emph{gravitational} energy contribution, \emph{nonlinear} in $\Lambda$ at order $r^5$ (plus an infinite number of higher order nonlinear terms). This suggests that it is naive to interpret the cosmological constant (times $c^4/(8\pi G)$) as simply a local ``matter'' energy volume density. It is that, locally, but it does not ``integrate'' trivially over a finite volume of space---there are nonlinear effects that also give rise to a ``gravitational'' energy interpretation. As alluded to in the Introduction, energy in general relativity is fundamentally non-local (or quasilocal) in nature, so the interpretation of $\Lambda$ as (proportional to) a ``local'' energy density of the gravitational vacuum can be made sense of only as an ``effective'' one---it is not truly (or only) local.

As a final consistency check, we can differentiate $\mathtt{E}_{\rm R}$ in (\ref{eq:RQFenergy}) with respect to $t$ and use the local matter energy conservation law ($\dot{\rho}=-3H(\rho+p)$) that follows from the two Friedmann equations to show that
\begin{equation}
\frac{{\rm d}\mathtt{E}_{\rm R}}{{\rm d}t}=\int_{\mathscr{S}_t}\, \bm{\epsilon}_{\mathscr{S}}^{\rm R}\, \left[-(\rho+p)\,\gamma^2\beta\right]\,,
\end{equation}
in agreement with (\ref{eq:MatterEnergyFlux}). 

We thus see that the rigid quasilocal frame energy conservation law in FLRW spacetimes is a highly non-trivial, nonlinear re-expression (and integration) of the \emph{local} matter energy conservation law (equivalent to $\nabla_a T^{ab}=0$) in a \emph{quasi}local form that includes both matter \emph{and} gravitational energy, in which the flat space vacuum energy plays a pivotal role. The co-moving quasilocal frame energy conservation law has a complementary form that involves only gravitational energy flux versus only matter energy flux. In short, we see that (\ref{eq:intconserve}) leads to a completely satisfactory energy conservation law in FLRW spacetimes, even for two very different sets of quasilocal observers, and which sheds some light on the subtleties of including gravitational energy in an energy conservation law.

It is worth comparing the above discussion with that of Afshar in Ref. \cite{afshar_quasilocal_2009}. In his equation~(43), Afshar evaluates Epp's ``invariant quasilocal energy''~\cite{epp_angular_2000} for co-moving observers and obtains the same result as in our (\ref{RQF_ver1}) or (\ref{eq:RQFenergy}), provided we make two changes: (1) we replace Afshar's expanding areal radius $ra(t)$ (suitable for his co-moving observers) with our constant areal radius $r$ (suitable for our RQF observers), and (2) we omit Afshar's reference energy subtraction. On point (1), the agreement between the two results makes sense since Epp's ``invariant quasilocal energy'' is like a ``rest mass'' in that it is boost gauge invariant, so the same for rigid observers as for co-moving observers instantaneously on the same sphere. This nicely emphasizes the point made after (\ref{eq:LorentzFactorRelation}) that the RQF quasilocal energy $\mathtt{E}_{\rm R}$ represents the energy as seen by ``stationary'' observers ``at rest'' in space, or ``at rest'' with respect to the system in question. Moreover, Afshar's calculation for the Brown-York quasilocal energy seen by co-moving observers---his equation (21), of course agrees with our (\ref{eq:ComovingEnergy}) for $\mathtt{E}_{\rm C}$, if again we omit Afshar's reference energy subtraction (and replace his $r$ with our $R$ notation for the co-moving radial coordinate). After his equation (21) Afshar suggests that, because this result makes no reference to the matter content of the model, that something seems to be ``missing.'' We argued above that there is, in fact, nothing missing. The matter content is clearly present in $\mathtt{E}_{\rm R}$, and insofar as $\mathtt{E}_{\rm C}$ is related to $\mathtt{E}_{\rm R}$ by the correct $\gamma$ factor, it is also present in $\mathtt{E}_{\rm C}$, except it appears now as purely gravitational energy (\emph{curved space} vacuum energy)---recall our discussion following (\ref{eq:ComovingEnergy}). In the previous paragraph we emphasized the pivotal role played by the \emph{flat space} vacuum energy; here we see the role of \emph{curved space} vacuum energy. Vacuum energy is clearly important, and we see no reason to include an ad hoc (flat space) vacuum reference energy subtraction, as is often done, including in Refs. \cite{afshar_quasilocal_2009} and \cite{epp_angular_2000}.

\section{Quasilocal energy of scalar cosmological perturbations}\label{sec:pert}

We turn our attention now to applying these ideas to a flat (${\sf k}=0$) FLRW metric with scalar perturbations in the Newtonian gauge and with a vanishing cosmological constant in the Einstein equation ($\Lambda =0$), and we aim to draw comparisons between our results (appropriately applied to a ``small locality'') and the more typical effective local treatment developed e.g. in Ref. \cite{abramo_energy-momentum_1997} and widely used in cosmology today.

To this end, we will construct and consider energy expressions for rigid quasilocal frames, i.e. for quasilocal observers which are ``at rest'' relative to each other (such that the spacetime physics is essentially
encoded in the energy-momentum boundary fluxes, and not in any relative motion of these observers on the boundary itself).

\subsection{Quasilocal conservation laws in general-relativistic perturbation theory}

We would like to begin by offering a brief general discussion on general-relativistic perturbation theory and perturbative gauge transformations in the context
of quasilocal conservation laws. A more detailed treatment of these
topics can be found in Sec. III of Ref. \cite{oltean_motion_2019} in the context of the gravitational
self-force problem. While the discussion in this subsection is not strictly speaking required to follow the work in the successive subsections of this section, and much of it is review of known results, our aim here is to remove as far as possible any conceptual ambiguity that the more technically-inclined reader may otherwise be disposed to question with regards to these topics. 

\begin{figure}
\begin{centering}
\includegraphics[scale=0.85]{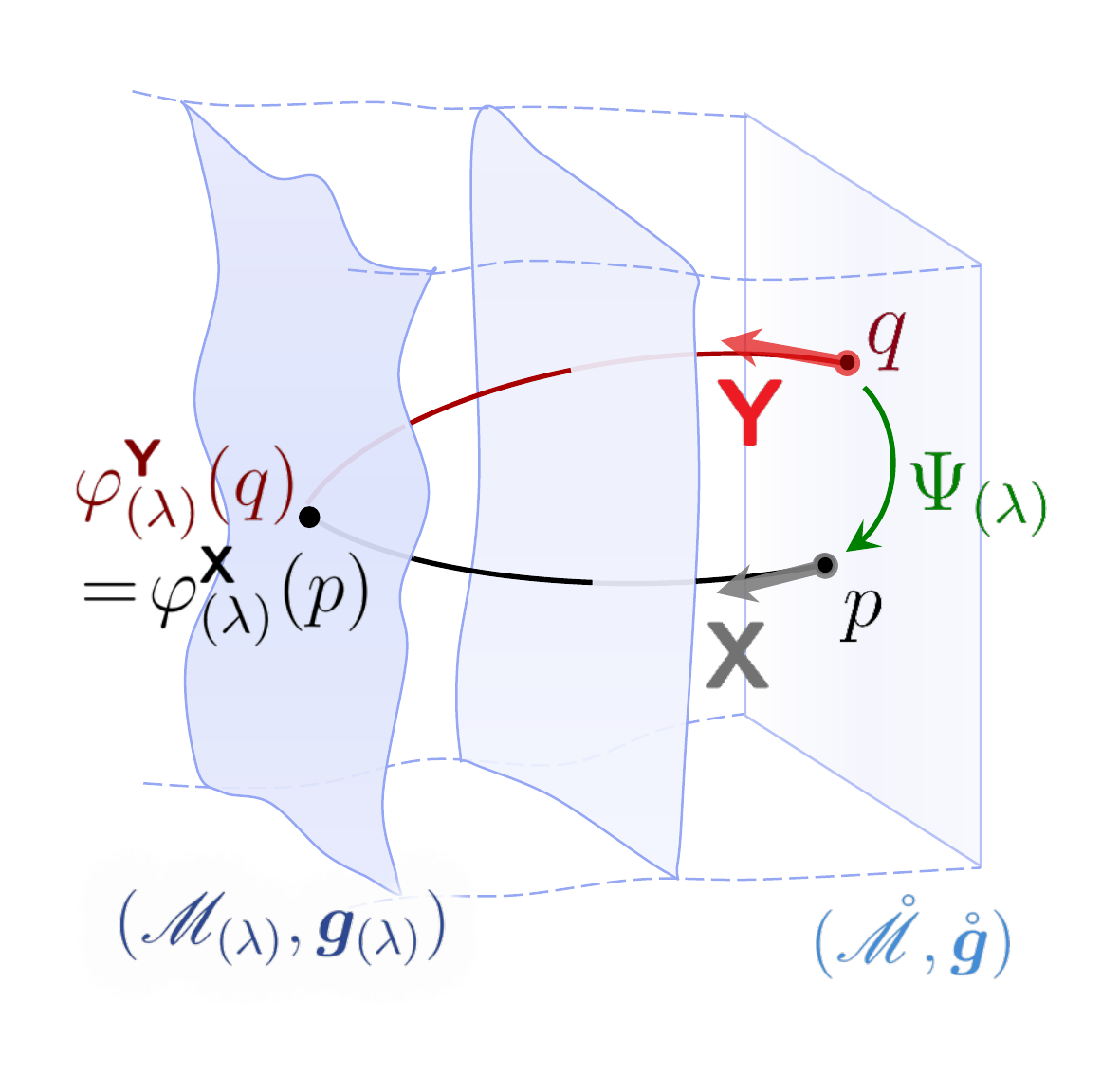} 
\par\end{centering}
\caption{Representation of a  one-parameter family of spacetimes  $\{\mathscr{M}_{(\lambda)}\}_{\lambda\geq0}$ used for perturbation theory. Each of the $\mathscr{M}_{(\lambda)}$ are depicted visually in $(1+1)$ dimensions,
as leaves of a (five-dimensional) product manifold $\mathscr{N}=\mathscr{M}_{(\lambda)}\times\mathbb{R}$, with the coordinate $\lambda\geq 0$ representing the perturbative expansion parameter.
A choice of a map (or gauge) $\varphi^{\bm{\mathsf{X}}}_{(\lambda)}:\mathring{\mathscr{M}}\rightarrow\mathscr{M}_{(\lambda)}$
gives us a way of identifying any point $p\in\mathring{\mathscr{M}}=\mathscr{M}_{(0)}$
on the background to one on some perturbed ($\lambda>0$)
spacetime, i.e. $p\protect\mapsto\varphi^{\bm{\mathsf{X}}}_{(\lambda)}(p)$. In particular, the collections of points $\{\varphi_{(\lambda)}^{\boldsymbol{\mathsf{X}}}(p)\}_{\lambda\geq0}$ are integral curves (e.g. the black curve in the figure) of a vector field $\boldsymbol{\mathsf{X}}\in T\mathscr{N}$, referred to as the gauge vetcor or also simply the gauge. A gauge transformation consists in choosing a different vector field in $T \mathscr{N}$, or equivalently a different associated diffeomorphism $\varphi_{(\lambda)}$, for identifying points between the background and the perturbed spacetimes. In this illustration, the point $p\in \mathring{\mathscr{M}}$ is mapped under the flow of $\bm{\mathsf{X}}$ to the same point in $\mathscr{M}_{(\lambda)}$ as is $q\in \mathring{\mathscr{M}}$ under the flow of $\bm{\mathsf{Y}}$ (for $p\neq q$ and $\bm{\mathsf{X}}\neq \bm{\mathsf{Y}}$). One thus has a gauge transformation on the background $q\mapsto \Psi_{(\lambda)}(q)=p$.}
\label{fig-gauge} 
\end{figure}

Let $\{(\mathscr{M}_{(\lambda)},g_{ab}^{(\lambda)})\}_{\lambda\geq0}$
denote any smooth one-parameter family of perturbed spacetimes in
the standard sense, such that $\lambda\geq0$ is the ``small'' perturbation
parameter. (See, e.g., Ref. \cite{bruni_perturbations_1997}.) Concordantly let $\mathscr{N}=\mathscr{M}_{(\lambda)}\times\mathbb{R}^{\geq}$
denote the product manifold produced by ``stacking'' together all
$\mathscr{M}_{(\lambda)}$, for all $\lambda\geq0$. For simplicity,
we denote by $\mathring{\mathscr{M}}=\mathscr{M}_{(0)}$ the background
manifold. See Fig. \ref{fig-gauge}.

A perturbative gauge is a map
\begin{align}
\varphi_{(\lambda)}^{\bm{\mathsf{X}}}:\mathring{\mathscr{M}}\rightarrow\, & \mathscr{M}_{(\lambda)}\\
p\mapsto\, & \varphi_{(\lambda)}^{\bm{\mathsf{X}}}(p)\,,
\end{align}
where $\bm{\mathsf{X}}\in T\mathscr{N}$ is the gauge vector (which may itself be referred to as the gauge),
i.e. a vector in the tangent space of the product manifold $\mathscr{N}=\mathscr{M}_{(\lambda)}\times\mathbb{R}^{\geq}$
the integral curves of which identify points between the background
$\mathring{\mathscr{M}}$ and all $\mathscr{M}_{(\lambda)}$ (for
all $\lambda>0$). See Fig \ref{fig-gauge}. Any different gauge vector $\bm{\mathsf{Y}}\in T\mathscr{N}$
will define (by its integral curves) a different identification via
a different map $\varphi_{(\lambda)}^{\bm{\mathsf{Y}}}:\mathring{\mathscr{M}}\rightarrow\mathscr{M}_{(\lambda)}$,
i.e. a different gauge. 

Equivalently as concerns computations on the background, and thus
for all practical purposes, a gauge can also be viewed as a local
choice of coordinates $\{x^{\alpha}\}$ on the background $\mathring{\mathscr{M}}$
to the desired perturbative order ($n$) at which one is working,
i.e. $x^{\alpha}=x_{(0)}^{\alpha}+\lambda x_{(1)}^{\alpha}+\ldots+\lambda^{n}x_{(n)}^{\alpha}+\mathcal{O}(\lambda^{n+1})$.
This is because in practice, we always ultimately resort to performing
calculations on the background in the form of tensor transports thereto
(under the map $\varphi_{(\lambda)}^{\bm{\mathsf{X}}}$, for a choice
of $\bm{\mathsf{X}}$) of equations in $\mathscr{M}_{(\lambda)}$
(for $\lambda>0$). (The choice of $\varphi_{(\lambda)}:\mathring{\mathscr{M}}\rightarrow\,\mathscr{M}_{(\lambda)}$
is often referred to as the ``active'' definition of the gauge,
and the associated choice of $\{x^{\alpha}\}$ in $\mathring{\mathscr{M}}$
as the ``passive'' definition of the same.) Concordantly, a gauge
transformation [from a gauge defined by a vector $\bm{\mathsf{Y}}\in T\mathscr{N}$
to one defined by $\bm{\mathsf{X}}\in T\mathscr{N}$, i.e. the background
map $\Psi_{(\lambda)}:\mathring{\mathscr{M}}\rightarrow\mathring{\mathscr{M}}$
defined by $\Psi_{(\lambda)}=(\varphi_{(\lambda)}^{\bm{\mathsf{X}}})^{-1}\circ\varphi_{(\lambda)}^{\bm{\mathsf{Y}}}=\varphi_{(-\lambda)}^{\bm{\mathsf{X}}}\circ\varphi_{(\lambda)}^{\bm{\mathsf{Y}}}$],
is in practice equivalent to a transformation (beginning at linear
order in $\lambda$),  $y^{\alpha}=x_{(0)}^{\alpha}+\sum_{j\geq1}\lambda^{j}y_{(j)}^{\alpha}\mapsto x^{\alpha}=x_{(0)}^{\alpha}+\sum_{j\geq1}\lambda^{j}x_{(j)}^{\alpha}$
(meaning, order-by-order beginning with first: $y_{(1)}^{\alpha}\mapsto x_{(1)}^{\alpha}$,
$\ldots$ , $y_{(n)}^{\alpha}\mapsto x_{(n)}^{\alpha}$ to the desired
order of computation $n$), of the local coordinates on the background
spacetime.

\begin{figure}
\begin{centering}
\includegraphics[scale=0.8]{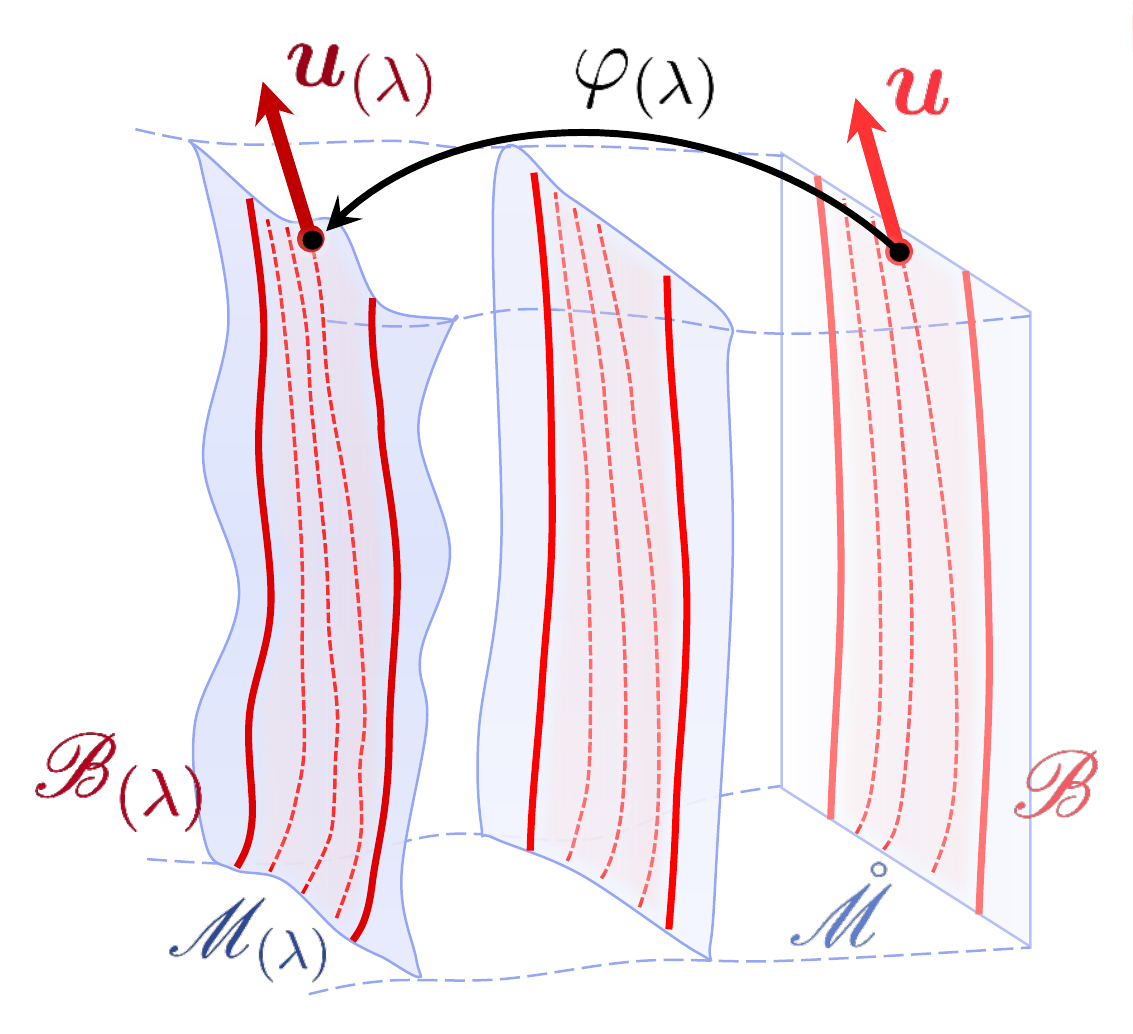} 
\par\end{centering}
\caption{Representation of a one-parameter family of quasilocal frames $\{(\mathscr{B}_{(\lambda)},u^{a}_{(\lambda)})\}_{\lambda\geq0}$
embedded correspondingly in a family of spacetimes $\{\mathscr{M}_{(\lambda)}\}_{\lambda\geq0}$.}
\label{fig-pertqf} 
\end{figure}

These ideas naturally extend to the construction of perturbed quasilocal conservation
laws. Let $\{(\mathscr{B}_{(\lambda)},u_{(\lambda)}^{a})\}_{\lambda\geq0}$
denote a one-parameter family of quasilocal frames such that $(\mathscr{B}_{(\lambda)},u_{(\lambda)}^{a})$
is embedded in $(\mathscr{M}_{(\lambda)},g_{ab}^{(\lambda)})$ for
each $\lambda\geq0$. See Fig. \ref{fig-pertqf}. 

Then let us consider here the quasilocal energy conservation
law (\ref{eq:intconserve}) (see Sec. III of Ref. \cite{oltean_motion_2019} for the equivalent discussion vis-à-vis the
quasilocal momentum conservation law) for $(\mathscr{B}_{(\lambda)},u_{(\lambda)}^{a})$
for $\lambda>0$,
\begin{equation}
\intop_{\mathscr{S}_{\textrm{f}}^{(\lambda)}-\mathscr{S}_{\textrm{i}}^{(\lambda)}}\bm{\epsilon}_{\mathscr{S}_{(\lambda)}}\,\mathcal{E}_{(\lambda)}=\intop_{\Delta\mathscr{B}_{(\lambda)}}\bm{\epsilon}_{\mathscr{B}_{(\lambda)}}\left[T_{(\lambda)}^{ab}u_{a}^{(\lambda)}n_{b}^{(\lambda)}-\tau_{(\lambda)}^{ab}\mathcal{D}_{a}^{(\lambda)}u_{b}^{(\lambda)}\right]\,,\label{eq:pertenergy}
\end{equation}
where $\mathcal{D}_{a}^{(\lambda)}$ is induced by the derivative
compatible with $g_{ab}^{(\lambda)}$ in $\mathscr{M}_{(\lambda)}$.
The idea now is to use the fact that for any diffeomorphism $f:\mathscr{U}\rightarrow\mathscr{V}$
between two manifolds $\mathscr{U}$ and $\mathscr{V}$ of the same
dimension, and for any form $\bm{\omega}$ of that dimension in $\mathscr{V}$,
we have that $\int_{\mathscr{V}}\bm{\omega}=\int_{\mathscr{U}}f^{*}\bm{\omega}$.
Thus the energy change in the perturbed spacetime [the LHS of (\ref{eq:pertenergy})] is
equal to the LHS of the equation below, which expresses a transported
energy change on the background; its RHS is a worldtube boundary flux
on the background, itself equal by the same token to the flux [the
RHS of (\ref{eq:pertenergy})] in the perturbed spacetime:
\begin{equation}
\intop_{\mathscr{S}_{\textrm{f}}-\mathscr{S}_{\textrm{i}}}\left[\left(\varphi_{(\lambda)}^{\bm{\mathsf{X}}}\right)^{*}\bm{\epsilon}_{\mathscr{S}_{(\lambda)}}\right]\,\mathcal{E}=\intop_{\Delta\mathscr{B}}\left[\left(\varphi_{(\lambda)}^{\bm{\mathsf{X}}}\right)^{*}\bm{\epsilon}_{\mathscr{B}_{(\lambda)}}\right]\left[T^{ab}u_{a}n_{b}-\tau^{ab}\mathcal{D}_{a}u_{b}\right]\,,
\end{equation}
where $\mathscr{S}=(\varphi_{(\lambda)}^{\bm{\mathsf{X}}})^{-1}(\mathscr{S}_{(\lambda)})\subset\mathring{\mathscr{M}}$,
similarly $\mathscr{B}=(\varphi_{(\lambda)}^{\bm{\mathsf{X}}})^{-1}(\mathscr{B}_{(\lambda)})\subset\mathring{\mathscr{M}}$,
and the quantities in the integrands involving no $(\lambda)$ sub-/super-scripts
are the tensor transports to the background spacetime (under the map
$\varphi_{(\lambda)}^{\bm{\mathsf{X}}}$) of the respective quantities
with such a sub-/super-script in $\mathscr{M}_{(\lambda)}$. In particular,
these become (in principle infinite order, truncated to the desired
order $n$ of computation) series in the perturbation parameter $\lambda$.

It should be clear that a practical computation (that is to say, on
the background spacetime) of any energy value, for example, makes
reference manifestly to a choice of some gauge vector $\bm{\mathsf{X}}$,
i.e. we have that the energy at a time slice $\mathscr{S}_{(\lambda)}$
is given by
\begin{equation}
\mathtt{E}=\int_{\mathscr{S}_{(\lambda)}}\bm{\epsilon}_{\mathscr{S}_{(\lambda)}}\left(\mathcal{E}_{(\lambda)}-\mathcal{P}_{(\lambda)}^{a}v_{a}^{(\lambda)}\right)=\int_{\mathscr{S}}\left[\left(\varphi_{(\lambda)}^{\bm{\mathsf{X}}}\right)^{*}\bm{\epsilon}_{\mathscr{S}_{(\lambda)}}\right]\,\left(\mathcal{E}^{\boldsymbol{\mathsf{X}}}-\mathcal{P}_{\boldsymbol{\mathsf{X}}}^{a}v_{a}^{\boldsymbol{\mathsf{X}}}\right)\,,
\end{equation}
with $\mathcal{E}^{\boldsymbol{\mathsf{X}}}=(\varphi_{(\lambda)}^{\bm{\mathsf{X}}})^{*}\mathcal{E}_{(\lambda)}$
and similarly for the other terms in the integrand of the final expression,
which is an intregral in the background.

We define rigid quasilocal frame-adapted coordinates $\{x^{\alpha}\}$
to be coordinates on $\mathring{\mathscr{M}}$ defined by a gauge
vector $\bm{\mathsf{R}}$ such that $\mathscr{S}=\mathbb{\mathbb{S}}^{2}$
and $(\varphi_{(\lambda)}^{\bm{\mathsf{R}}})^{*}\bm{\epsilon}_{\mathscr{S}_{(\lambda)}}=\bm{\epsilon}_{\mathbb{S}^{2}}$
in these coordinates. Indeed, it was shown in Sec. IV B of Ref. \cite{oltean_motion_2019} that, given
a rigid quasilocal frame $(\mathscr{B}_{(\lambda)},u_{(\lambda)}^{a})$
in $\mathscr{M}_{(\lambda)}$, its inverse image $(\mathscr{B},u^{a})$
in $\mathring{\mathscr{M}}$ (i.e. the background congruence with
four-velocity $u^{a}=(\varphi_{(\lambda)})^{*}u_{(\lambda)}^{a}$)
in general deviates from quasilocal rigidity in the background (with
respect to the background metric) beginning at linear order in $\lambda$.
Setting order-by-order these deviations to zero could be regarded
as defining the gauge $\boldsymbol{\mathsf{R}}$ (up to the order
of computation).

In practice, one may not have given $g_{ab}^{\boldsymbol{\mathsf{R}}}=(\varphi_{(\lambda)}^{\bm{\mathsf{R}}})^{*}g_{ab}^{(\lambda)}$
to carry out computations in principle in a rigid quasilocal 
frame-adapted gauge. Instead, suppose we have $g_{ab}^{\boldsymbol{\mathsf{Y}}}$
in any gauge $\boldsymbol{\mathsf{Y}}$ with associated coordinates
$\{y^{\alpha}\}$, and to simplify notation let us denote in general the
surface density $\varrho^{\boldsymbol{\mathsf{Y}}}=\mathcal{E}^{\boldsymbol{\mathsf{Y}}}-\mathcal{P}_{\boldsymbol{\mathsf{Y}}}^{a}v_{a}^{\boldsymbol{\mathsf{Y}}}$ for any $\boldsymbol{\mathsf{Y}}$.
The energy $\mathtt{E}$ can thus be expressed as an integral of this
surface density over a closed two-surface with volume form $(\varphi_{(\lambda)}^{\bm{\mathsf{Y}}})^{*}\bm{\epsilon}_{\mathscr{S}_{(\lambda)}}$
which is not in general that of an exact two-sphere (to the perturbative
order of computation). Rather than attempting to perform such an integral
(which may not be possible analytically), it may in practice prove
to be far easier to obtain the transformation from the $\boldsymbol{\mathsf{Y}}$
to the $\boldsymbol{\mathsf{R}}$ gauge, i.e. the map $y^{a}\mapsto x^{\alpha}$,
and to compute the integral in the latter:
\begin{equation}
\mathtt{E}=\int_{\mathscr{S}}\left[\left(\varphi_{(\lambda)}^{\bm{\mathsf{Y}}}\right)^{*}\bm{\epsilon}_{\mathscr{S}_{(\lambda)}}\right]\,\varrho^{\boldsymbol{\mathsf{Y}}}=\int_{\mathbb{S}^{2}}\bm{\epsilon}_{\mathbb{S}^{2}}\varrho^{\boldsymbol{\mathsf{R}}}\,,
\end{equation}
which is thus an integral over an exact two-sphere. The latter may prove far easier (perhaps even analytically possible at all) to carry out vis-à-vis the former, i.e. an integral over a non-spherical closed two-surface $\mathscr{S}$.

Note that the $\boldsymbol{\mathsf{Y}}$ gauge integral (over $\mathscr{S}$, which
in general is not $\mathbb{S}^{2}$) 
evaluates, naturally, to a result in the $\{y^{\alpha}\}$ coordinates, and the
$\boldsymbol{\mathsf{R}}$ gauge integral (over $\mathbb{S}^{2}$) to one in $\{x^{\alpha}\}$. Thus, if
one is ultimately interested in the results of a computation in the
former, one may compute the latter, $\int_{\mathbb{S}^{2}}\bm{\epsilon}_{\mathbb{S}^{2}}\varrho^{\boldsymbol{\mathsf{R}}}$
(as a function of $x^{\alpha}$) and transform the result back (to
a function of $y^{\alpha}$) by obtaining and applying the inverse
transformation, $x^{\alpha}\mapsto y^{\alpha}$.

Specifically turning now to the problem of the perturbed quasilocal
energy for cosmological spacetimes, the first question to consider
is the choice of gauge $\boldsymbol{\mathsf{Y}}$. In particular,
we will compute $\mathtt{E}^{\boldsymbol{\mathsf{Y}}}$ (and hence,
an effective local energy volume density determined therefrom) for
a choice of $\boldsymbol{\mathsf{Y}}$ that will permit us a comparison
of our results with previous computations in the literature, that is to say, computations performed
in the same gauge. There is nothing forcing us to perform such a computation of the energy
in a certain gauge rather than any other gauge beyond the non-sensicality
of comparing energies (or energy densities) computed in different
gauges.

We will perform this calculation for $\boldsymbol{\mathsf{Y}}$ being
the Newtonian gauge, for which previous computations of effective
local energy densities are available with which to compare our results. We could
certainly perform the same for any other $\boldsymbol{\mathsf{X}}\neq\boldsymbol{\mathsf{Y}}$,
however we are not aware of a computation in the literature of an
effective local energy density of cosmological perturbations in a
gauge that is not the Newtonian one and thus to which we could compare
our result computed in such a gauge $\boldsymbol{\mathsf{X}}$. Moreover,
checking the consistency of such a comparison requires knowing the
transformation from the $\boldsymbol{\mathsf{Y}}$ to the $\boldsymbol{\mathsf{X}}$
gauge to second order, which in general cannot be expressed analytically; rather, it requires the numerical solution
to a set of coupled PDEs determining the components of the gauge transformation
vector. (See e.g. Sec. 5 of Ref. \cite{bruni_perturbations_1997} where the case of the transformation
from the Newtonian to the Poisson gauge, a generalization of the Newtonian gauge, is treated and the corresponding
set of PDEs is presented.) Therefore such an analysis, for any $\boldsymbol{\mathsf{X}}\neq\boldsymbol{\mathsf{Y}}$,
is beyond the scope of the present work.

\subsection{Setup: scalar cosmological perturbations in the Newtonian gauge}

We begin with the perturbed flat FLRW metric $g_{ab}$ in the Newtonian
gauge, in physical time $T$ and Cartesian spatial coordinates $\{X^{I}\}$
given collectively by\footnote{\setstretch{0.9}We remind the reader that Greek indices are used instead of
Latin ones to indicate a particular coordinate choice instead of abstract
index notation.} 
\begin{equation}
\{X^{\bar{\alpha}}\}=\{T,X^{I}\}_{I=1}^{3}\,,
\end{equation}
and denoting the metric perturbation as $\psi(X^{\bar{\alpha}})$,
\begin{equation}
g_{\bar{\alpha}\bar{\beta}}{\rm d}X^{\bar{\alpha}}{\rm d}X^{\bar{\beta}}=-\left(1+2\lambda\psi\right){\rm d}T^{2}+a^{2}\left(T\right)\left(1-2\lambda\psi\right)\delta_{IJ}{\rm d}X^{I}{\rm d}X^{J}+\mathcal{O}(\lambda^{2})\,,\label{eq:pert_FLRW}
\end{equation}
where we are ignoring quadratic ($\mathcal{O}(\lambda^{2})$) metric
perturbations\footnote{\setstretch{0.9}The quadratic metric perturbations should in fact be included in an exact calculation following our approach as a matter of consistency, and this inclusion can readily be accommodated. We are nevertheless ignoring these terms here for simplicity since in applications, these---just like the linear perturbations---are ``averaged out'' to zero \cite{abramo_energy-momentum_1997}, i.e. $\langle g_{\bar{\alpha}\bar{\beta}}^{(2)}\rangle_{\Sigma_{(0)}}=0$, just as $\langle g_{\bar{\alpha}\bar{\beta}}^{(1)}\rangle_{\Sigma_{(0)}}=0$ (while in general, $\langle(g_{\bar{\alpha}\bar{\beta}}^{(1)})^{2}\rangle_{\Sigma_{(0)}}\neq0$).}. 

We assume that this is sourced by a scalar matter field,
described by a usual matter stress-energy-momentum tensor
\begin{equation}
T_{ab}=\nabla_{a}\varphi\nabla_{b}\varphi-\frac{1}{2}g_{ab}\left(\nabla_{c}\varphi\nabla^{c}\varphi+2V\left[\varphi\right]\right)\,,\label{eq:T_ab_scalar}
\end{equation}
for which we assume a perturbative expansion with only a time-dependent
background,
\begin{equation}
\varphi(X^{\bar{\alpha}})=\varphi_{0}(T)+\lambda\Delta(X^{\bar{\alpha}})+\mathcal{O}(\lambda^{2})\,.
\end{equation}

It is known that, at zeroth order in $\lambda$, the (respectively,
time-time and space-space) Einstein equations are the usual Friedmann
equations\footnote{\setstretch{0.9}The space-space
equation (\ref{eq:background_space-space}) is written after division by $a^{2}$.},
\begin{align}
3H^{2}=\, & 8\pi\left(\frac{1}{2}\dot{\varphi}_{0}^{2}+V_{0}\right)\,,\label{eq:background_time-time}\\
-\left(2\dot{H}+3H^{2}\right)=\, & 8\pi\left(\frac{1}{2}\dot{\varphi}_{0}^{2}-V_{0}\right)\,,\label{eq:background_space-space}
\end{align}
where overdot indicates a derivative with respect to $T$, and we denote
$H=\dot{a}/a$ and $V_{0}=V[\varphi_{0}]$. (Note that the second equation can
also be rewritten using the first as $\dot{H}+3H^{2}=8\pi V_{0}$).
At first order, we have the (respectively, time-time, time-space and
and space-space) Einstein equations\footnote{\setstretch{0.9}The space-space equation (\ref{eq:pert_space-space}) is written after some LHS-RHS cancellation
between Einstein and matter tensor components and division by $a^{2}$.}
\begin{align}
-6H\dot{\psi}+\frac{2}{a^{2}}\delta^{IJ}\partial_{I}\partial_{J}\psi\, & =8\pi\left(\dot{\varphi}_{0}\dot{\Delta}+\frac{\delta V_{0}}{\delta\varphi_{0}}\Delta+2V_{0}\psi\right)\,,\label{eq:pert_time-time}\\
\partial_{I}\left(2\dot{\psi}+2H\psi\right)\, & =8\pi\partial_{I}\left(\dot{\varphi}_{0}\Delta\right)\,,\\
2\ddot{\psi}+8H\dot{\psi}+16\pi V_{0}\psi\, & =8\pi\left(\dot{\varphi}_{0}\dot{\Delta}-\frac{\delta V_{0}}{\delta\varphi_{0}}\Delta\right)\,.\label{eq:pert_space-space}
\end{align}

Additionally, the Einstein equations for $g_{\bar{\alpha}\bar{\beta}}$
are supplemented with the matter field equations for the scalar field
$\varphi$ described by the matter tensor $T_{ab}$ above {[}Eq. (\ref{eq:T_ab_scalar}){]},
in particular the Klein-Gordon equation $\nabla^{a}\nabla_{a}\varphi-\delta V/\delta\varphi=0$.
In the $\{X^{\bar{\alpha}}\}$ coordinates, at zeroth order, this
is
\begin{equation}
-\ddot{\varphi}_{0}-3H\dot{\varphi}_{0}-\frac{\delta V_{0}}{\delta\varphi_{0}}=0\,,
\end{equation}
and at linear perturbative order it is
\begin{equation}
-\ddot{\Delta}-3H\dot{\Delta}+\left(\frac{1}{a^{2}}\delta^{IJ}\partial_{I}\partial_{J}-\frac{\delta^{2}V_{0}}{\delta\varphi_{0}^{2}}\right)\Delta+4\dot{\varphi}_{0}\dot{\psi}-2\psi\frac{\delta V_{0}}{\delta\varphi_{0}}=0\,.
\end{equation}

\subsection{Construction of rigid quasilocal frames}

We now follow the same general method as in the previous section to
construct here a quasilocal frame\footnote{\setstretch{0.9}We drop the ``R'' sub-/super-script as it is understood that we are working here and for the remainder of this section only with rigid quasilocal frames.} $(\mathscr{B},u^{a})$: that is,
by performing a change of coordinates from the Newtonian-gauge coordinates
$\{X^{\bar{\alpha}}\}=\{T,X^{I}\}$ to a set of adapted spherical
coordinates 
\begin{equation}
\{x^{\alpha}\}=\{t,r,\theta,\phi\}\,.
\end{equation}
The latter are said to be adapted to the quasilocal observers in the
sense that $\mathscr{B}=\{r={\rm const}.\}$, and such that the congruence
four-velocity is given, in these coordinates, by
\begin{equation}
u^{\alpha}=\frac{1}{N}\delta^{\alpha}\,_{t},
\end{equation}
where $N$ is the lapse of $g_{\alpha\beta}$. By orthogonality we
thus have that the normal vector to $\mathscr{B}$ is $n^{\alpha}=g^{\alpha r}/\sqrt{g^{rr}}$,
and from these we can proceed to compute all necessary geometrical
quantities defined in Sec. \ref{sec:qf}. Different choices of such a transformation
$\{X^{\bar{\alpha}}\}\mapsto\{x^{\alpha}\}$ can thus be regarded
as corresponding to different types of quasilocal frames. 

We wish to compute the second-order energy of the linear metric perturbation
$\psi$. Thus, for this we would like to construct a quasilocal frame
that is rigid up to second perturbative order inclusive (modulo linear
terms in quadratic ($\mathcal{O}(\lambda^{2})$) metric perturbations,
which we are ignoring for this calculation)\footnote{\setstretch{0.9}We remind the reader that the only reason we do this here is for computational
simplicity, with a view towards comparing our results with those of
Ref. \cite{abramo_energy-momentum_1997} where the quadratic metric perturbations are ``averaged out''
to zero. See again footnote 14. If desired, we could certainly add
the quadratic perturbations to the perturbative expansion of the metric
in our calculation here, and investigate the conditions for its formal
averaging following the methods of e.g. Ref. \cite{Gasperini_2009}. This would be interesting
in terms of developing further the connection between our work here
and the averaging methods used in cosmology, but regardless of how
the latter are carried out, the averaging of $g_{\alpha\beta}^{(2)}$
cannot introduce any new terms involving the linear perturbation variables
$g_{\alpha\beta}^{(1)}$. Thus it is consistent for us to neglect
$g_{\alpha\beta}^{(2)}$ in our calculation here, as any results at
second perturbative order (e.g. for the energy) would receive
additional terms only involving $g_{\alpha\beta}^{(2)}$ under a different
averaging procedure, possibly.

~

~

}.

We begin by considering
a coordinate transformation of the general from
\begin{equation}
\begin{cases}
T\left(x^{\alpha}\right) & =t+\lambda f_{(1)}\left(x^{\alpha}\right)+\lambda^{2}f_{(2)}\left(x^{\alpha}\right)\,,\\
X^{I}\left(x^{\alpha}\right) & =\left(1+\lambda F_{(1)}\left(x^{\alpha}\right)+\lambda^{2}F_{(2)}\left(x^{\alpha}\right)\right)\frac{r}{a(t)}r^{I}\,,
\end{cases}\label{eq:coord_trans_lambda2}
\end{equation}
where $r^{I}=\left(\sin\theta\cos\phi,\sin\theta\sin\phi,\cos\theta\right)$
are the standard direction cosines of a radial unit vector in $\mathbb{R}^{3}$
and $f_{(1)}$, $f_{(2)}$, $F_{(1)}$ and $F_{(2)}$ are all functions of
the $\{x^{\alpha}\}$ coordinates. We choose these functions so as
to achieve an exact two-sphere metric induced on each constant-time
slice in these coordinates, i.e. such that
\begin{equation}
\sigma_{\alpha\beta}=g_{\alpha\beta}-n_{\alpha}n_{\beta}+u_{\alpha}u_{\beta}=r^{2}{\rm diag}\left(0,0,1,\sin^{2}\theta\right)+\mathcal{O}(\lambda^{3})\,.\label{eq:sigma_alpha_beta}
\end{equation}

Proceeding thus, i.e. computing $\sigma_{\alpha\beta}$ in the coordinates
$\{x^{\alpha}\}$ given by the general transformation (\ref{eq:coord_trans_lambda2})
and then setting it equal to (\ref{eq:sigma_alpha_beta}), we find
that we can satisfy the latter (boundary rigidity) condition with
the choice 
\begin{equation}
\begin{cases}
f_{(1)}=-\mathcal{H}^{-1}\psi\,, & f_{(2)}=-\frac{1}{2}\mathcal{H}^{-3}\left(\mathcal{H}'+2\mathcal{H}^{2}\right)\psi^{2}\,,\\
F_{(1)}=\psi\,, & F_{(2)}=\frac{3}{2}\psi^{2}\,,
\end{cases}\label{eq:coord_trans_functions}
\end{equation}
where $'={\rm d}/{\rm d}t$ and $\mathcal{H}=a'/a$.\footnote{\setstretch{0.9}Note that our use
here of the prime ($'$) and calligraphic Hubble parameter ($\cal{H}$) symbols should not be confused with their more typical usage in the cosmology
literature where they are usually related to a different time coordinate,
namely conformal time. In our case, we emphasize that $'$ indicates
a derivative with respect to the time coordinate $t$ adapted to quasilocal
observers.}

\subsection{Energy of cosmological perturbations}

In the choice of coordinates $\{x^{\alpha}\}$ described above {[}Eq.
(\ref{eq:coord_trans_lambda2}) with $f_{(1)}$, $f_{(2)}$, $F_{(1)}$
and $F_{(2)}$ given by Eq. (\ref{eq:coord_trans_functions}){]}, we
compute the Brown-York tensor $\Pi_{\alpha\beta}$ from which we get
the quasilocal energy (boundary) density: 
\begin{align}
\mathcal{E}=\, & -\frac{1}{8\pi}\Pi_{\alpha\beta}u^{\alpha}u^{\beta}\\
=\, & \mathcal{E}_{(0)}+\lambda\mathcal{E}_{(1)}+\lambda^{2}\mathcal{E}_{(2)}+\mathcal{O}(\lambda^{3})\,,
\end{align}
where 
\begin{align}
\mathcal{E}_{(0)}\left(x^{\alpha}\right)=\, & \underset{\,}{-\frac{\sqrt{1-\mathcal{H}^{2}r^{2}}}{4\pi r}\,,}\\
\mathcal{E}_{(1)}\left(x^{\alpha}\right)=\, & \underset{\,}{-\frac{(\mathcal{H}'+\mathcal{H}^{2})r}{4\pi\sqrt{1-\mathcal{H}^{2}r^{2}}}\psi\,,}\\
\mathcal{E}_{(2)}\left(x^{\alpha}\right)=\, & \underset{\,}{\frac{(\frac{\mathcal{H}''}{\mathcal{H}}+2\mathcal{H}'+4\mathcal{H}^{2})r+\frac{(\mathcal{H}'+\mathcal{H}^{2})^{2}}{(1-\mathcal{H}^{2}r^{2})}r^{3}}{8\pi\sqrt{1-\mathcal{H}^{2}r^{2}}}\psi^{2}\,.}
\end{align}
We will also need the quasilocal momentum, which we compute to be:
\begin{align}
\mathcal{P}^{\alpha}=\, & \frac{1}{8\pi}\sigma^{\alpha\beta}u^{\gamma}\Pi_{\beta\gamma}\\
=\, & \mathcal{P}_{(0)}^{\alpha}+\lambda\mathcal{P}_{(1)}^{\alpha}+\lambda^{2}\mathcal{P}_{(2)}^{\alpha}+\mathcal{O}(\lambda^{3})\,,
\end{align}
where 
\begin{align}
\mathcal{P}_{(0)}^{\alpha}\left(x^{\alpha}\right)=\, & \underset{\,}{0\,,}\\
\mathcal{P}_{(1)}^{\alpha}\left(x^{\alpha}\right)=\, & \underset{\,}{-\frac{\delta^{\alpha}\,_{\mathfrak{i}}\mathcal{H}\sigma^{\mathfrak{ij}}\partial_{\mathfrak{j}}\psi}{8\pi r(1-\mathcal{H}^{2}r^{2})}\,,}\\
\mathcal{P}_{(2)}^{\alpha}\left(x^{\alpha}\right)=\, & \underset{\,}{-\frac{\delta^{\alpha}\,_{t}(\nabla_{\mathbb{S}^{2}}\psi)^{2}+\delta^{\alpha}\,_{\mathfrak{i}}[\psi'-(\mathcal{H}+\frac{\mathcal{H}'}{\mathcal{H}}+\frac{2\mathcal{H}}{1-\mathcal{H}^{2}r^{2}})\psi]\sigma^{\mathfrak{ij}}\partial_{\mathfrak{j}}\psi}{8\pi r(1-\mathcal{H}^{2}r^{2})}\,.}
\end{align}
These are exact results for any rigid quasilocal frame $(\mathscr{B},u^{a})$
in this spacetime, where we recognize $\mathcal{E}_{(0)}$ as the background
energy already obtained and discussed in the previous section.

Now let us consider the total energy, 
\begin{equation}
\mathtt{E}=\int_{\mathscr{S}}\bm{\epsilon}_{\mathscr{S}}^{\,}\,\left(\mathcal{E}-\mathcal{P}^{a}v_{a}\right)=\mathtt{E}_{(0)}+\lambda\mathtt{E}_{(1)}+\lambda^{2}\mathtt{E}_{(2)}+\mathcal{O}(\lambda^{3})\,,
\end{equation}
where in these coordinates we have $\mathscr{S}=\mathbb{S}_{r}^{2}$
and $\bm{\epsilon}_{\mathscr{S}}=\bm{\epsilon}_{\mathbb{S}_{r}^{2}}=r^{2}\bm{\epsilon}_{\mathbb{S}^{2}}$.
Moreover we can use the fact that $v_{\mathfrak{i}}=-u_{\mathfrak{i}}$
to compute
\begin{equation}
\int_{\mathscr{S}}\bm{\epsilon}_{\mathscr{S}}^{\,}\,\left(-\mathcal{P}^{a}v_{a}\right)=-\lambda^{2}\frac{r}{8\pi \sqrt{1-\mathcal{H}^{2}r^{2}}}\int_{\mathbb{S}_{r}^{2}}\bm{\epsilon}_{\mathbb{S}^{2}}(\nabla_{\mathbb{S}^{2}}\psi)^{2}+\mathcal{O}(\lambda^{3})\,.
\end{equation}

It is interesting now to consider $\mathtt{E}$ in a small-radius
expansion. Vis-à-vis the above ($-\mathcal{P}^{a}v_{a}$) term, one can easily show and use the fact that
\begin{equation}
\int_{\mathbb{S}_{r}^{2}}\bm{\epsilon}_{\mathbb{S}^{2}}(\nabla_{\mathbb{S}^{2}}\psi)^{2}=r^{2}\frac{8\pi}{3}(\nabla_{\mathbb{R}^{3}}\psi)^{2}+\mathcal{O}(r^{3})\,,
\end{equation}
where the Cartesian spatial three-gradient is understood to be evaluated
at $r=0$.

The linear energy ($\mathtt{E}_{(1)}$) is usually regarded in applications as physically
uninteresting, as $\psi$ is typically ``averaged out'' to zero over all
of three-space \cite{abramo_energy-momentum_1997}, i.e. $\langle\psi\rangle_{\Sigma_{(0)}}=0$. The second-order
energy, after substitution of the background Einstein equations (Friedmann
equations), simplifies to 
\begin{equation}
\mathtt{E}_{(2)}=\left(\frac{4\pi}{3}r^{3}\right)\rho_{(2)}+\mathcal{O}(r^{4})\,,
\end{equation}
where we have defined an effective local energy volume density 
\begin{equation}
\rho_{(2)}\left(x^{\alpha}\right)=3\left(\frac{2\mathcal{H}^{2}}{\pi}-4V_{0}+\frac{\varphi_{0}'}{\mathcal{H}}\frac{\delta V_{0}}{\delta\varphi_{0}}\right)\psi^{2}-\frac{1}{4\pi}(\nabla_{\mathbb{R}^{3}}\psi)^{2}\,,
\end{equation}
where $\psi$ as well as its spatial gradient are both now understood to be evaluated at $r=0$. Thus, rigid quasilocal observers only see a ``mass-type'' (metric perturbation squared) and Cartesian three-gradient squared correction to the effective local energy density away from FLRW. These arise, respectively, from the quasilocal energy surface density $\mathcal{E}$ and from the quasilocal momentum shift projection ($-\mathcal{P}^{a}v_{a}$).

It is interesting now to study this result if we transform back to the original
$\{X^{\bar{\alpha}}\}$ coordinate system (i.e. the Newtonian perturbative gauge in which the metric (\ref{eq:pert_FLRW}) and field equations (\ref{eq:background_time-time})-(\ref{eq:pert_space-space}) were written), in order to make a connection
with the work of Ref. \cite{abramo_energy-momentum_1997} and so that we
can more conveniently use, if we so wish, the Einstein
equations (\ref{eq:background_time-time})-(\ref{eq:pert_space-space}) in order to make simplifying substitutions. For this transformation, we also need the zeroth and
linear order effective local energy volume densities in the $\{x^{\alpha}\}$
coordinates (defined in the same way as $\rho_{(2)}$ at their respective
orders), 
\begin{align}
\rho_{(0)}\left(x^{\alpha}\right)=\, & \frac{3\mathcal{H}^{2}}{8\pi}\,,\\
\rho_{(1)}\left(x^{\alpha}\right)=\, & \left(\frac{3\mathcal{H}^{2}}{2\pi}-6V_{0}\right)\psi\,,
\end{align}
which upon performing the transformation $\{x^{\alpha}\}\mapsto\{X^{\bar{\alpha}}\}$
will contribute $\mathcal{O}(\lambda^{2})$ terms to $\rho_{(2)}$.
To do this, we simply need to set $\mathcal{H}=JH$ where $H=\dot{a}/a$
(with overdot denoting a derivative with respect to $T$) as before,
and we define $J$ to be the time coordinate Jacobian, $J=\partial T/\partial t$.
We must compute the latter from the transformation {[}Eq. (\ref{eq:coord_trans_lambda2})
and (\ref{eq:coord_trans_functions}){]} and re-express it in terms
of the $\{X^{\bar{\alpha}}\}$ coordinates. We find: 
\begin{equation}
J=\frac{\partial T}{\partial t}=1+\lambda J_{(1)}+\lambda^{2}J_{(2)}+\mathcal{O}(\lambda^{3})\,,
\end{equation}
where 
\begin{align}
J_{(1)}(X^{\bar{\alpha}})=\, & H^{-2}\left(-H\dot{\psi}+\dot{H}\psi\right)\,,\\
J_{(2)}(X^{\bar{\alpha}})=\, & \tfrac{1}{2}H^{-4}\left[-2H^{2}\psi\ddot{\psi}+2H\left(\dot{H}-2H^{2}\right)\psi\dot{\psi}+\left(H\ddot{H}-\dot{H}^{2}+2H^{2}\dot{H}\right)\psi^{2}\right]\,.
\end{align}
Using this, we get $\rho_{(2)}$ in the original ($\{X^{\bar{\alpha}}\}$) coordinates:
\begin{equation}
\rho_{(2)}(X^{\bar{\alpha}})=\frac{3}{8\pi}\dot{\psi}^{2}-\frac{3}{4\pi}\psi\ddot{\psi}-\frac{9H}{2\pi}\psi\dot{\psi}+\left(\frac{3H^{2}}{2\pi}+6\frac{\dot{\varphi}_{0}}{H}\frac{\delta V_{0}}{\delta\varphi_{0}}\right)\psi^{2}-\frac{1}{4\pi a^{2}}(\partial_{I}\psi)^{2}\,.
\end{equation}
This expresses the total---matter plus gravitational---energy (effective local density) still strictly in terms of the gravitational perturbation ($\psi$) without a direct appearance of the matter perturbation ($\Delta$). In order to see how the latter plays a role, and also to eliminate the second time derivative term which is a priori unusual in an energy expression, we can substitute $\ddot{\psi}$ from the (dynamical) space-space
first-order Einstein equation (\ref{eq:pert_space-space}) to get:
\begin{equation}
\rho_{(2)}=\frac{3}{8\pi}\dot{\psi}^{2}-\frac{3H}{2\pi}\psi\dot{\psi}+\left(\frac{3H^{2}}{2\pi}+6V_{0}+6\frac{\dot{\varphi}_{0}}{H}\frac{\delta V_{0}}{\delta\varphi_{0}}\right)\psi^{2}+3\left(-\dot{\varphi}_{0}\dot{\Delta}+\frac{\delta V_{0}}{\delta\varphi_{0}}\Delta\right)\psi-\frac{1}{4\pi a^{2}}(\partial_{I}\psi)^{2}\,.\label{eq:rho_2_no-ddot}
\end{equation}

We note that in our expression for $\rho_{(2)}$ above [Eq. (\ref{eq:rho_2_no-ddot})], the first two terms
coincide with the negative of the first two terms obtained in Eq. (65) of Ref. \cite{abramo_energy-momentum_1997} for the averaged effective local
energy density of scalar cosmological perturbations, i.e. the $t_{00}$ in our notation in the introduction, Eq. (\ref{eq:t_ab}) [called ``$\tau_{00}$'' in Ref. \cite{abramo_energy-momentum_1997}, not the same as the Brown-York tensor $\tau_{ab}$ used in this paper]. We will return to comment more upon this sign discrepancy in the following subsection.


\subsection{Specialization to some time-only dependent cases of interest}

Let us further study these results in the simplified case where the perturbations are only time-dependent.

\subsubsection{Vanishing potential}

In the case that $V_0=0$, the time-time Einstein equation (\ref{eq:pert_time-time}) tells us that $\dot{\psi}$ is proportional $\dot{\Delta}$ with the factor dependent on background quantities. (In particular, $-3H\dot{\psi}=4\pi\dot{\varphi}_{0}\dot{\Delta}$.) Hence a kinetic energy term for the matter perturbation can be brought to appear explicitly by substituting $\dot{\psi}$ from this Einstein equation into (\ref{eq:rho_2_no-ddot}). This yields:
\begin{equation}
\rho_{(2)}=\frac{1}{2}\dot{\Delta}^{2}+\frac{3H}{4\pi}\psi\dot{\psi}+\frac{3H^{2}}{2\pi}\psi^{2}\,.
\end{equation}

Indeed, in this simplified case, we have that $\frac{3}{8\pi}\dot{\psi}^{2}=\frac{1}{2}\dot{\Delta}^{2}$ exactly (with the equality holding up to spatial dependence and potential terms in the general case), thanks to the first-order and background time-time Einstein equations. Now, in our notation, the $t_{00}$ (``$\tau_{00}$'' therein) of Ref. \cite{abramo_energy-momentum_1997} (to be compared, in principle, to our $\rho_{(2)}$), in this reduced case reads: 
\begin{equation}
t_{00}=-\frac{3}{8\pi}\langle\dot{\psi}^{2}\rangle_{\Sigma_{(0)}}+\frac{3H}{2\pi}\langle\psi\dot{\psi}\rangle_{\Sigma_{(0)}}+\frac{1}{2}\langle\dot{\Delta}^{2}\rangle_{\Sigma_{(0)}}\,,\label{eq:t_00}
\end{equation}
and so would be left without any explicit kinetic energy terms upon substituting $\frac{3}{8\pi}\dot{\psi}^{2}=\frac{1}{2}\dot{\Delta}^{2}$, i.e. we are only left with $t_{00}=\frac{3H}{2\pi}\langle\psi\dot{\psi}\rangle_{\Sigma_{(0)}}$ in this case, where it does not seem possible to use further field equation substitutions to produce a kinetic-type (perturbation time derivative squared) term. Thus we conjecture that the factors appearing in front of the (gravitational and matter) perturbation kinetic terms in (\ref{eq:t_00}) [Eq. (65) of Ref. \cite{abramo_energy-momentum_1997} in our notation and in this reduced case] are not consistent, but rather, the true, \textit{total} kinetic energy of the perturbations (the time derivative squared term in the effective density), based on our results, is either the kinetic energy of $\psi$ only (with a factor of $+3/8\pi$),
\begin{equation}
\rho_{\textrm{KE}}^{(\psi)}=\frac{3}{8\pi}\dot{\psi}^{2}\,,
\end{equation} 
or (equivalently thanks to the Einstein equation) the kinetic energy of $\Delta$ only (with the usual factor of $+1/2$),
\begin{equation}
\rho_{\textrm{KE}}^{(\Delta)}=\frac{1}{2}\dot{\Delta}^{2}\,,
\end{equation}
or any combination thereof permitted by further Einstein equation substitutions.

\subsubsection{Slow-roll approximation}

In the slow-roll approximation, relevant for inflation, we have $V_{0}\gg\frac{1}{2}\dot{\varphi}_{0}^{2}$
so that $H^{2}\approx\frac{8\pi}{3}V_{0}$ and $\frac{\delta V_{0}}{\delta\varphi_{0}}\approx-H\dot{\varphi}_{0}$.
Furthermore approximating $\dot{\psi}\approx0$, we thus get an effective
density:
\begin{equation}
\rho_{(2)}\approx4V_{0}\psi^{2}\,,
\end{equation}
which coincides with the result in Eq. (74) of \cite{abramo_energy-momentum_1997}
up to sign and a potential second functional derivative term. We conjecture that this sign discrepancy might be connected to the sign discrepancy we have obtained with the general metric perturbation kinetic term, and so this disagreement warrants further investigation. Physically, this is important because the sign of this $\rho_{(2)}$ (positive as in our result, or negative as in the result of Ref. \cite{abramo_energy-momentum_1997}) determines the overall effect of perturbations on inflation (either accelerating it further or, respectively, counteracting it).

\section{Conclusions}\label{sec:concl}

\subsection{Summary and discussion of results}

In this paper, we have offered an initial investigation of quasilocal
(matter plus gravitational) energy-momentum, specifically using the
Brown-York tensor, in cosmological solutions of general relativity.
In particular, we have computed and investigated the quasilocal energy
of homogeneous isotropic (FLRW) spacetimes and of scalar cosmological
perturbations.

Gravitational energy-momentum is fundamentally quasilocal in nature. Thus, the total energy-momentum (of gravity plus matter) for any physical system must also be fundamentally quasilocal, as must be its conservation law. In the local limit, such a quasilocal conservation law must reduce to $\nabla^{a} T_{ab}=0$ (a local conservation law for matter alone), where the quantity $T_{ab}$, which serves as the local source of the gravitational field, could be thought of as an emergent limit of the more fundamental quasilocal (matter plus gravity) energy-momentum. In trying to apply the principle of energy-momentum conservation in cosmology, researchers usually begin with the local notion of matter energy-momentum and try to include some form of effective local gravitational energy-momentum. We are advocating that a better approach is to use a quasilocal (total) energy-momentum conservation law as the starting point.

As we have seen, the quasilocal energy is purely geometrical in character,
in other words the total (gravitational plus matter) energy within
any spacetime region is given purely in terms of the intrinsic and extrinsic geometry of the spacetime boundary of that region.
The explicit appearance
of matter terms in energy expressions can then be achieved by local substitutions
of the Einstein equation, as we have seen explicitly with our application
of these definitions in this work to cosmological spacetimes. The
lesson is that while the Einstein equation expresses the local dynamics
between the (local) stress-energy-momentum of matter fields and the
(local) gravitational field, and while a local definition of the energy-momentum
of the latter is fundamentally incompatible with the precepts of the
theory, a purely geometrical quasilocal (boundary) density of stress-energy-momentum
is generally capable of capturing a meaningful notion of this concept
as applied to the \textit{total} physical field, i.e. of both the matter and
the gravitational field. In other words, the manifestation of the
stress-energy-momentum of non-gravitational (i.e. matter) fields as
local ought to be regarded as an effective phenomenon due to spacetime
geometry acting as a physical field itself (one locally sourced thereby),
and thus fundamentally emergent from a (purely geometrical) quasilocal
notion.

This point of view is relevant when regarded in the wider context
that much of the history of applications of general relativity involving
notions of ``gravitational energy-momentum'' has for reasons of
understandable expediency---and frequently, in appropriate approximations,
leading thus to much practical success---attempted to treat this
as an effective local phenomenon, similarly to how we are used to
thinking about matter pre-relativistically. However such approaches
will always be fundamentally limited by the extent to which the approximations
assumed in the problem permit or not the existence of a (sufficiently) mathematically
well-defined and physically sensible effective ``gravitational energy-momentum''
notion of such a (local) sort. Rather, the essential lesson of the equivalence
principle is that the situation is other way around: matter stress-energy-momentum
is an ``effectively local'' phenomenon (and attributed this interpretation
as the local source of the gravitational field), and emergent from
the energy of the total physical field, including that of gravity,
which is fundamentally quasilocal. Thus, beginning with a quasilocal
point of view on stress-energy-momentum in problems of application
involving such notions as attributed to both gravitational and non-gravitational
fields may shed conceptual light and offer technical pathways forward
beyond what is available solely within an effective local perspective.

In Sec. \ref{sec:flrw}, we have applied the Brown-York definition of the quasilocal
(gravitational plus matter) stress-energy-momentum along with the
notion of quasilocal frames to homogeneous isotropic (FLRW) spacetimes
and considered their energy conservation laws. We have shown that
these are able to recover the usual effective local manifestation
of the energy density and pressure of matter as well as of a cosmological
constant term in the Einstein equation. The latter is often interpreted
as the (effective local) ``energy/pressure of the gravitational vacuum'',
often dubbed ``dark energy'', and much dispute has arisen from attempts
to reconcile such a notion with the fundamentally local treatment
of energy in matter (non-gravitational) theories such as quantum field theory. However,
passing to a quasilocal treatment---which fundamentally accounts for
both matter and gravity---may help to shed fresh perspectives on
this issue: perhaps the relevant question to ask is not how gravitational
energy-momentum can be ``fitted within'' a local perspective (as we are used to taking with non-gravitational theories),
yet strictly in an effective way, so as to make sense of phenomena
such as the observed accelerated expansion along with the matter
content of our Universe, but instead, how the matter plus gravitational
energy can be described together, quasilocally, to make sense fundamentally
of such phenmomena. 

In Sec. \ref{sec:pert}, we have applied these quasilocal notions to computing
the energy of scalar cosmological perturbations. Historically this
has been approached a priori via the effective local method, thus
requiring various assumptions to be introduced---with different ones
taken by different authors---to formulate the problem. This has led
to often divergent conclusions on the question of the back-reaction
of the perturbations upon the background metric, relevant especially
for early Universe cosmology. As we have shown here, the Brown-York
quasilocal energy (applied to a rigid quasilocal frame) can recover sensible expressions for
the effective second-order (in perturbation theory) local
energy density of the total (gravitational plus matter) scalar cosmological
perturbations, comparable to the standard results obtained via ``averaging'' arguments \cite{abramo_energy-momentum_1997}. We emphasize however that our approach is \textit{exact}, without the need to introduce any averaging assumptions from the beginning, and indeed for this reason we do not expect a priori a direct term-by-term comparison between our (exact, unaveraged) effective local energy density and the (averaged) result of Ref. \cite{abramo_energy-momentum_1997}. In fact, we conjecture that the kinetic energies in the latter are not consistent (in particular, the total kinetic energy of the perturbations therein seems to vanish in a time-only dependent and vanishing potential case), and that based on our results, the true, total (matter plus gravitational) kinetic energy is given either as that of the metric perturbation ($\psi$), $\rho_{\textrm{KE}}^{(\psi)}=\frac{3}{8\pi}\dot{\psi}^{2}$, or equivalently, that of the matter perturbation ($\Delta$), $\rho_{\textrm{KE}}^{(\Delta)}=\frac{1}{2}\dot{\Delta}^{2}$ (with $\rho_{\textrm{KE}}^{(\psi)}=\rho_{\textrm{KE}}^{(\Delta)}$ modulo spatial dependence and potential terms).

\subsection{Outlook}

This work merely begins to scratch the surface of the potential utility
of quasilocal stress-energy-momentum definitions and conservation
laws for cosmology. The ideas outlined here can be extended in a wide
variety of directions.

While in this work we have employed the Brown-York definition for
the quasilocal stress-energy-momentum, it would be interesting to
investigate further and compare cosmological results using other quasilocal
definitions, as was done e.g. in Refs. \cite{afshar_quasilocal_2009} and \cite{faraoni_newtonian_2015,faraoni_turnaround_2015,faraoni_beyond_2017} using the Epp and Hawking-Hayward definitions respectively. Moreover, here we have worked within general relativity,
and so one may consider appropriately generalized definitions of quasilocal
tensors to investigate cosmological solutions in modified gravitational
theories. (A simple approach would be to retain the same basic Brown-York
definition $\tau_{ab}\propto\delta S_{\textrm{total}}/\delta g^{ab}$
but applied to any modified theory $S_{\textrm{total}}$, not necessarily
general relativity minimally coupled to matter.) 

Furthermore, here we have investigated the FLRW metric with no perturbations
and only with scalar metric perturbations in the Newtonian gauge.
The analysis can be extended to include all---scalar, vector and tensor---modes of the linear metric perturbations. Our formalism naturally permits the treatment of higher-order perturbations as well, if desired. 

Moreover, it would be very interesting to investigate these results further in specific
situations such as inflation and the problem of cosmological back-reactions in the early Universe. The manifest advantage of our approach is that it is exact, without the need to introduce from the start any simplifying assumptions or ``averaging'' procedures. This may help to provide a more fundamental perspective and to clarify the current disputes in the cosmology community on this issue arising from different assumptions and averaging procedures being used for effective (local) approaches. Quasilocal methods may also provide insights into the flow of both matter \emph{and} gravitational energy-momentum during large-scale structure formation.

Finally, yet another very interesting direction into which these concepts could be taken is the study of gravitational entropy in a cosmological context. For example, a more recent idea similar to quasilocal frames, dubbed ``gravitational screens'' \cite{freidel_gravitational_2015,freidel_non-equilibrium_2015}, motivated more from thermodynamic considerations, was proposed and developed. A detailed comparison between our formalism and Refs. \cite{freidel_gravitational_2015,freidel_non-equilibrium_2015} remains lacking, but would be very interesting to undertake especially from the point of view of defining and working with a general notion of entropy. An application of this to cosmology could also make contact with the work of Refs. \cite{faraoni_cosmological_2011,uzun_quasilocal_2015}. 

\section*{Acknowledgements}

We thank Robert H. Brandenberger and Carlos F. Sopuerta for useful discussions. HBM is supported by Ferdowsi University of Mashhad.

%

\end{document}